\documentclass[%
pra,
reprint,
amsfonts,
amsmath,
amssymb,
aps,
]{revtex4-2}

\usepackage{graphicx}
\usepackage{bm}
\usepackage{bbm}
\usepackage{physics}
\usepackage{tikz}
\usepackage{environ}
\usepackage{pgfplots}
\usepackage{makecell}
\usepackage{pifont}
\newcommand{\cmark}{\ding{51}}%
\newcommand{\xmark}{\ding{55}}%
\usepackage{hyperref}
\usepackage{url}
\usetikzlibrary{decorations.pathmorphing, positioning, shapes,arrows,arrows.meta}
\usetikzlibrary{shapes.geometric}
\usetikzlibrary{fadings}
\usetikzlibrary{external}
\usepgfplotslibrary{fillbetween}
\tikzfading 
[
  name=fade out,
  inner color=transparent!0,
  outer color=transparent!100
]

\usepackage{natbib}

\usepackage[final]{changes}
\usepackage{xr}
\begin{document}
\title{Moment-based superresolution: Formalism and applications}
\author{Giacomo Sorelli} 
\affiliation{Laboratoire Kastler Brossel, Sorbonne Universit\'e, ENS-Universit\'e PSL, CNRS, Collège de France, 4  Place Jussieu, F-75252 Paris, France}
\author{Manuel Gessner}
\affiliation{Laboratoire Kastler Brossel, Sorbonne Universit\'e, ENS-Universit\'e PSL, CNRS, Collège de France, 4  Place Jussieu, F-75252 Paris, France}
\author{Mattia Walschaers} 
\affiliation{Laboratoire Kastler Brossel, Sorbonne Universit\'e, ENS-Universit\'e PSL, CNRS, Collège de France, 4  Place Jussieu, F-75252 Paris, France}
\author{Nicolas Treps}
\affiliation{Laboratoire Kastler Brossel, Sorbonne Universit\'e, ENS-Universit\'e PSL, CNRS, Collège de France, 4  Place Jussieu, F-75252 Paris, France}

\date{\today}

\begin{abstract}
Sensitivity limits are usually determined using the Cram\'er-Rao bound. 
Recently this approach has been used to obtain the ultimate resolution limit for the estimation of the separation between two incoherent point sources.
However, methods that saturate these resolution limits, usually require the full measurement statistics, which can be challenging to access.
In this work, we introduce a simple superresolution protocol to estimate the separation between two thermal sources which relies only on the average value of a single accessible observable. 
We show how optimal observables for this technique may be constructed for arbitrary thermal sources, and
we study their sensitivities when one has access to spatially resolved intensity measurements (direct imaging) and photon counting after spatial mode demultiplexing.
For demultiplexing, our method is optimal, i.e. it saturates the quantum Cram\'er-Rao bound.
We also investigate the impact of noise on the optimal observables, their measurement sensitivity and on the scaling with the number of detected photons of the smallest resolvable separation.
For low signals in the image plane, we demonstrate that our method saturates the Cram\'er-Rao bound even in the presence of noise.
\end{abstract}

\maketitle
\section{Introduction}
Resolving small angular separations through an optical imaging system is an important problem both in microscopy and in astronomy.
The most traditional imaging technique is a spatially resolved intensity measurement, also known as direct imaging. 
The resolution of this approach, as pointed out already by Abbe \cite{Abbe} and Rayleigh \cite{Rayleigh} at the end of the nineteenth century, is limited by diffraction and noise. 
However, for a large enough signal-to-noise ratio, the diffraction limit is not fundamental \cite{goodman2015statistical}, and can be overcome by superresolution techniques based on fluorescence microscopy \cite{Hell:94, Klar8206, Betzig1642}, homodyne measurements \cite{Hsu_2004, Delaubert:06, PinelPRA2012} or intensity measurements in an appropriate spatial mode basis \cite{Helstrom73,Tsang2019}.

Recently, superresolution imaging was analysed from the point of view of quantum metrology 
\cite{helstrom1976, BraunsteinCaves1994, holevo2011probabilistic,Paris2009,
GiovannettiLoydMaccone,Luca_Augusto_review}, and the Fisher information was used to determine how well the separation between two incoherent sources can be resolved.
In this framework, the diffraction limit manifests itself through the vanishing of the direct-imaging Fisher information for small source separations \cite{Tsang_PRX}.
On the other hand, the quantum Fisher information, i.e. the Fisher information optimized over all possible measurements in the image plane, stays constant for small distances \cite{Tsang_PRX}, leaving room for superresolution.
Furthermore, it was shown \cite{Tsang_PRX} that the ultimate resolution limit given by the quantum Fisher information can be approached by photon counting after spatial-mode demultiplexing. 
Several experiments \added{used spatial light modulators to} implemented a simplified version of this demultiplexing technique, which distinguishes only between the fundamental and the first excited modes \cite{Paur:16, Tang:16, Yang:16, Tham:2017}\replaced{ On the other hand}{,
while} distance estimation from the demultiplexing of multiple spatial modes\added{, realized using multi-plane light conversion \cite{Morizur:10},} was only recently reported \cite{Boucher:20}.

In general, to reach the Cram\'er-Rao bound,  e.g. via maximum likelihood estimation, requires to measure the full photon counting statistics, which can be practically challenging. 
Here, we demonstrate that this is not necessary in superresolution imaging, where the ultimate resolution can be obtained using a moment-based estimation technique, that requires to measure only the average value of a single measurement observable.
For this estimation technique, we identify the optimal observables when different measurements, such as spatially resolved intensity measurements (direct imaging) or photon counting after spatial mode demultiplexing, are available.

In particular, we focus on the estimation of the transverse separation between two thermal sources of arbitrary, and different brightnesses, and
we determine the sensitivity that can be reached with our optimized observables, and consequently the minimal resolvable distance between the sources.
For demultiplexing measurements, we construct the optimal observables also in presence of experimental imperfections such as misalignment, measurement crosstalk, and detector noise. 
Therefore, our results are directly relevant for practical applications.
Even in the presence of noise, for low brightnesses of the sources, we prove that our approach is sufficient to saturate the Cram\'er-Rao bound. 
Finally, for arbitrary brightnesses of the sources, we demonstrate that our optimized demultiplexing measurement allows to reach the quantum Crámer-Rao bound \cite{Nair_2016, LupoPirandola}, in the noiseless scenario, if sufficiently many modes are measured.

The paper is structured as follows:
First, in Sec.~\ref{Sec:state_image}, we present our model for thermal sources in the image plane of an imaging system.
After recalling the method of moments for parameter estimation in Sec.~\ref{Sec:MethodOfMoments}, in Sec.~\ref{Sec:construction}, we use it to construct the optimal observable to estimate the source separation via spatial mode demultiplexing, while in Sec.~\ref{Sec:Direct_imaging}, we employ it to bound the sensitivity of ideal direct imaging.
Detailed studies of the performances of our moment-based approach for ideal and noisy demultiplexing are presented in Sec.~\ref{Sec:Moments_ideal}, and Sec.~\ref{Sec:Additional_res} respectively.
In Sec.~\ref{Sec:min_res_dist}, we discuss the smallest source separation which is resolvable by the different measurement techniques.
Section \ref{Sec:conclusion} concludes our work.

\section{Thermal states in the image plane}
\label{Sec:state_image}
We want to estimate the transverse distance between two point thermal sources located at positions $\pm {\bf r}_0$, with ${\bf r}_0 = (d\cos\theta/2, d\sin\theta/2)$.
The sources emit a total mean photon number equal to $2N$ in the spatial modes associated with the field operators $\hat{s}_1$ and $\hat{s}_2$. 
We denote with $\hat{\rho}_a(N)$ a thermal state with mean photon number $N$ in the mode associated with the field operator $\hat{a}$.
Accordingly, the sources are described by the quantum state $\hat{\rho}_0 = \hat{\rho}_{s_1}[(1-\gamma)N] \otimes \hat{\rho}_{s_2}[(1+\gamma)N]$ where $-1 < \gamma < 1$ takes into account the possibly different (but finite) temperatures of the two sources.
Such a state is described by the density matrix 
\begin{equation}
\hat{\rho}_0 = \int d^2 \alpha_{1} d^2\alpha_{2} P_{s_1,s_2} (\alpha_{1}, \alpha_{2})\ket{\alpha_1,\alpha_2}\bra{\alpha_1,\alpha_2},
\label{rhos1s2}
\end{equation}
where $\ket{\alpha_{1/2}}$ are coherent states of the field operators $\hat{s}_{1/2}$, and $P_{s_1,s_2} (\alpha_{1}, \alpha_{2}) = P_{s_1}(\alpha_{1})P_{s_2}(\alpha_{2})$ is the Glauber-Sudarshan $P-$function, with 
\begin{equation}
P_{s_{1/2}}(\alpha_{1/2}) =\frac{1}{\pi N (1\mp\gamma)} e^{ - |\alpha_{1/2}|^2 / (1\mp\gamma)N}.
\label{P_term}
\end{equation}

\begin{figure}
\includegraphics[width =\columnwidth]{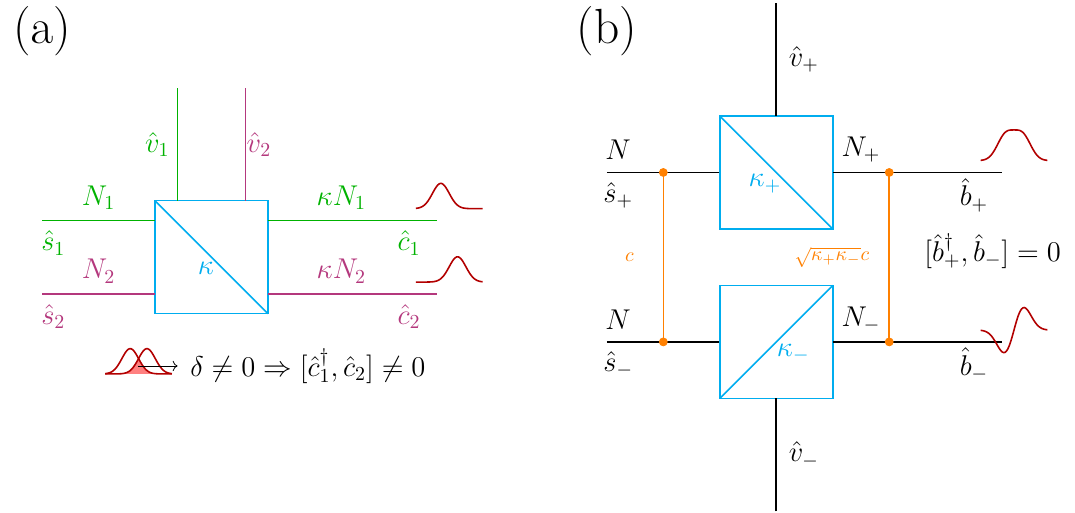}
\caption{Equivalent beam-splitter models for the propagation of thermal states through a diffraction-limited imaging system. (a) The source's modes $\hat{s}_{1/2}$ are populated with two thermal states with photon numbers $N(1\pm \gamma)$, and are mixed with \added{the non-orthogonal} vacuum modes $\hat{v}_{1/2}$ on \replaced{a beam splitter with}{two beam splitters with equal} transmissivity $\kappa$, resulting in the non-orthogonal image modes $\hat{c}_{1/2}$. (b) The symmetric and antisymmetric modes $\hat{s}_{\pm}$ have equal mean photon number $N$, but are classically correlated with phase-insensitive correlations $\langle \hat{s}_\pm^\dagger \hat{s}_\mp \rangle =  \gamma N$.
The modes $\hat{s}_\pm$ are mixed with \added{ the orthogonal} vacuum modes $\hat{v}_\pm$ on two beam splitters with transmissivities $\kappa_\pm$, resulting in the orthogonal image modes $\hat{b}_\pm$.}
\label{Fig:sources}
\end{figure}

The evolution of the field operators through a diffraction-limited imaging system, with transmissivity $\kappa$, is given by \cite{Shapiro_rev,LupoPirandola}
\begin{subequations}
\begin{align}
\hat{c}_1 &= \sqrt{\kappa}\hat{s}_1 + \sqrt{1-\kappa}\hat{v}_1,\\
\hat{c}_2 &= \sqrt{\kappa}\hat{s}_2 + \sqrt{1-\kappa}\hat{v}_2,
\end{align}
\label{map12}
\end{subequations}
where $\hat{c}_{1/2}$ are the field operators associated with the images $u_0({\bf r} \pm {\bf r}_0)$ of the two sources, with $u_0({\bf r})$ the point spread function (PSF) of the imaging system, which we assume to be real up to a global phase.
The field operators $\hat{v}_{1/2}$ are associated with auxiliary modes, that are in the vacuum state. 
This beam-splitter model for the propagation through an imaging system is illustrated in Fig.~\ref{Fig:sources} (a).

The modes $u_0({\bf r} \pm {\bf r}_0)$ are non-orthogonal, and therefore the operators $\hat{c}_1^\dagger$ and $\hat{c}_2$ do not commute (see Fig.~\ref{Fig:sources} (a)). 
As a consequence, these modes cannot be used to properly represent the quantum state of the sources in the image plane. 
To obviate this problem, we introduce the orthonormal image modes
\begin{equation}
u_{\pm}({\bf r}) = \frac{u_0({\bf r} + {\bf r}_0) \pm u_0({\bf r} - {\bf r}_0)}{\sqrt{2(1\pm \delta)}},
\label{upm}
\end{equation}
where $\delta$ represents the overlap between the source images
\begin{equation}
\delta = \int d^2 {\bf r} u^*_0({\bf r} + {\bf r}_0)u_0({\bf r} - {\bf r}_0).
\label{delta}
\end{equation}
The relation between the field operators $\hat{b}_{\pm}$ associated to the modes $u_{\pm}({\bf r})$ and the field operators $\hat{s}_\pm = (\hat{s}_1\pm \hat{s}_2)/\sqrt{2}$ in the object plane can be obtained from the sum and difference of Eqs.~\eqref{map12} \cite{LupoPirandola}
\begin{equation}
\hat{b}_{\pm} = \sqrt{\kappa_\pm}\hat{s}_\pm + \sqrt{1-\kappa_\pm}\hat{v}_\pm,
\label{mapping}
\end{equation}
with $\hat{v}_\pm$ associated with auxiliary modes, that are in the vacuum state, and $\kappa_\pm = \kappa(1\pm\delta)$.

We now use Eq.~\eqref{mapping} to propagate the quantum state $\hat{\rho}_0$ of the sources to the image plane.
First, we note that the transformation to the modes $\hat{s}_{\pm}$ can be interpreted as a $50:50$ beam splitter
\begin{equation}
\begin{pmatrix}
\hat{s}_+\\
\hat{s}_-
\end{pmatrix} =
\frac{1}{\sqrt{2}}\begin{pmatrix}
1 & 1\\
1 & -1
\end{pmatrix}
\begin{pmatrix}
\hat{s}_1\\
\hat{s}_2
\end{pmatrix}.
\label{50:50}
\end{equation}
According to Eq.~\eqref{50:50}, the coherent states $\ket{\alpha_1}\ket{\alpha_2}$ correspond to the coherent states $\ket{\frac{\alpha_1 + \alpha_2}{\sqrt{2}}}\ket{\frac{\alpha_1 - \alpha_2}{\sqrt{2}}} \equiv \ket{\alpha_+}\ket{\alpha_-}$ of the field operators $\hat{s}_\pm$ (see Fig.~\ref{Fig:sources} (b) for a schematic illustration).
The quantum state of the sources can therefore be rewritten in terms of the latter coherent states as
\begin{equation}
\hat{\rho}_0 = \int d^2 \alpha_{+} d^2 \alpha_{-} P_{s_+,s_-} (\alpha_{+}, \alpha_{-})\ket{\alpha_+,\alpha_-}\bra{\alpha_+,\alpha_-},
\label{rhosSsA}
\end{equation}
where $P_{s_+,s_-} (\alpha_{+}, \alpha_{-}) = P_{s_1,s_2} \left(\frac{\alpha_{+}+\alpha_-}{\sqrt{2}}, \frac{\alpha_{+}-\alpha_-}{\sqrt{2}}\right)$.
Accordingly, the modes $u_{\pm}({\bf r})$ have both mean photon number $\langle \hat{s}^\dagger_\pm \hat{s}_\pm \rangle = N$, and the photon number imbalance $\gamma$ appears in the classical, phase-insensitive correlations $\langle \hat{s}^\dagger_\pm \hat{s}_\mp \rangle = N\gamma$ (see orange lines in Fig. \ref{Fig:sources}. (b)).
Going through the imaging system according to Eq.~\eqref{mapping}, we have $\ket{\alpha_+,\alpha_-} \to \ket{\sqrt{\kappa_+} \alpha_+,\sqrt{\kappa_-} \alpha_-} \equiv \ket{\beta_+, \beta_-}$, with 
$\ket{\beta_{\pm}}$ coherent states of the field operators $\hat{b}_{\pm}$.
We can therefore write the quantum state in the image plane as
\begin{equation}
\hat{\rho}(d,\theta) = \int d^2 \beta_+ d^2 \beta_- P_{b_+,b_-}(\beta_+, \beta_-)\ket{\beta_+, \beta_-} \bra{\beta_+, \beta_-},
\label{rho_image}
\end{equation}
with 
\begin{equation}
P_{b_+,b_-}(\beta_+, \beta_-) = \frac{1}{\kappa_+\kappa_-}P_{s_+,s_-}\left(\frac{\beta_+}{\sqrt{\kappa_+}}, \frac{\beta_-}{\sqrt{\kappa_-}}\right).
\label{P_image_bb}
\end{equation}
Combining Eqs.~\eqref{P_image_bb} and \eqref{P_term}, we can write the $P$-function as
\begin{align}
P_{b_+,b_-}(\beta_+,\beta_-) = \frac{1}{\pi^2 \det V} e^{-{\bm \beta}^\dagger V {\bm \beta}},
\label{P_image}
\end{align}
where we have defined ${\bm \beta} = (\beta_+, \beta_-)^T$ and 
\begin{equation}
V  = \begin{pmatrix}
N_+ & \gamma\sqrt{N_+N_-}\\
\gamma\sqrt{N_+N_-} & N_-
\end{pmatrix},
\end{equation}
with $N_{\pm}= N \kappa_\pm$.
For equally bright sources ($\gamma = 0$), the off-diagonal elements of the matrix $V$ vanish, and Eq.~\eqref{P_image} reduces to the product of two Gaussian functions corresponding to $ \hat{\rho}(\theta, d) = \hat{\rho}_{b_+}(N_+) \otimes \hat{\rho}_{b_-}(N_-)$ as reported in \cite{sorelli2021optimal}.

Finally, the sources in the image plane are described by Eqs.~\eqref{rho_image} and \eqref{P_image}, with the information on parameter $d$ contained in the shape of the modes $u_\pm({\bf r})$ and the mean photon numbers $N_\pm$.

\section{The method of moments}
\label{Sec:MethodOfMoments}
We estimate the distance $d$ between the two sources with the method of moments \cite{Luca_Augusto_review, GessnerPRL2019}.
Following this approach, given an observable $\hat{X}$, an estimator $\tilde{d}$ for the parameter $d$ is extracted from the sample mean $\bar{x}_\mu = \sum_{i=1}^\mu x_i/\mu$ of $\mu$ independent measurements of $\hat{X}$.
The distance estimator is obtained by comparing the sample mean $\bar{x}_\mu$ with a calibration curve given by the expectation value $\langle \hat{X} \rangle = {\rm tr} [\hat{X} \hat{\rho}(d,\theta) ]$ of the measurement operator $\hat{X}$ as a function of the source separation $d$,  which could be known either from theory, or from a previous calibration experiment.

When sufficiently many measurements are performed $(\mu \gg 1)$, it follows from the central limit theorem that $\bar{x}_\mu$ is normally distributed with mean value $\langle \hat{X} \rangle$ and variance $(\Delta \hat{X})^2/\mu$, with $(\Delta \hat{X})^2 = \langle \hat{X}^2 \rangle - \langle \hat{X} \rangle^2$.
Accordingly, we choose as a separation estimator, the parameter value $\tilde{d}$ at which $\langle \hat{X} \rangle$ equates the measurements mean $\bar{x}_\mu$.
The associated estimation error is given by $(\Delta d)^2 =\chi^2[d,\theta,\hat{X}]/\mu$, with 
\begin{equation}
\chi^2[d,\theta,\hat{X}] = \frac{(\Delta \hat{X})^2}{\left( \frac{\partial \langle \hat{X} \rangle}{\partial d}\right)^2}.
\end{equation}
The quantity $\chi^2[d,\theta,\hat{X}]$ determines the sensitivity of the method-of-moments estimation strategy for the quantum state $\rho(d,\theta)$, and the observable $\hat{X}$. 
It obeys the following chain of inequalities
\begin{equation}
\chi^{-2}[d, \theta, \hat{X}] \leq \mathcal{F}[d, \theta, \hat{X}] \leq \mathcal{F}_Q\left[d, \theta\right],
\label{chain}
\end{equation}
where the second is saturable by an optimal operator $\hat{X}$ \cite{Luca_Augusto_review, BraunsteinCaves1994, kholevo1974, frowis2015, GessnerPRL2019}.
Here, $\mathcal{F}[d, \theta, \hat{X}]$ is the Fisher information that bounds the achievable sensitivity when estimating $d$ from measurements of $\hat{X}$ according to the Crámer-Rao lower bound $(\Delta d)^2  \geq (\mu \mathcal{F}[d, \theta, \hat{X}])^{-1}$ \cite{helstrom1976, BraunsteinCaves1994}.
Finally,  $\mathcal{F}_Q\left[d, \theta\right] = \max_{\hat{X}} \left[ \mathcal{F}[d, \theta, \hat{X}] \right]$ is the quantum Fisher information which defines the ultimate metrological sensitivity \cite{BraunsteinCaves1994}.

In practical situations, one does not have access to all possible measurement observables.
Therefore, we now assume that we can measure only a finite number $K$ of observables $\hat{X}_k$, as well as their linear combinations $\hat{X}_{\bf \tilde{m}} = \tilde{\bf m}\cdot \hat{\bf X}$, with $\hat{\bf X} = (\hat{X}_1, \dots, \hat{X}_K)^T$ and $\tilde{\bf m}$ the measurement coefficients vector.
Under these assumptions, it is possible to perform an analytical optimization over all possible linear combinations. 
Such an optimization yields the measurement sensitivity \cite{GessnerPRL2019}
\begin{align}
M[d,\theta, \hat{\bf X}] &= \max_{\tilde{m}} \chi^{-2}[d, \theta, \hat{X}_{\tilde{\bf m}}] \nonumber \\ &= {\bf D}[d,\theta,\hat{\bf X}]^T \Gamma[d,\theta,\hat{\bf X}]^{-1}{\bf D}[d,\theta,\hat{\bf X}],
\label{M}
\end{align}
where $\Gamma[d,\theta, \hat{\bf X}]$ is the  covariance matrix, whose elements are given by $\Gamma_{k,l} [d,\theta, \hat{\bf X}] = \langle \hat{X}_k \hat{X}_l \rangle -  X_k X_l$, while $D[d,\theta, \hat{\bf X}] = \frac{\partial X_k}{\partial d}$ is the derivative vector, where we labelled as $X_k = \langle \hat{X}_k \rangle$ the expectation value of $\hat{X}_k$. 
The sensitivity given by Eq.~\eqref{M} is achieved by the measurement coefficients $\tilde{\bf m} = {\bf m}$ \cite{GessnerPRL2019}, with 
\begin{equation}
{\bf m} = \eta \Gamma^{-1}[d, \theta, \hat{\bf X}] D[d, \theta,\hat{\bf X}],
\label{m}
\end{equation}
where $\eta$ is a normalization constant. 
\begin{figure}
\includegraphics{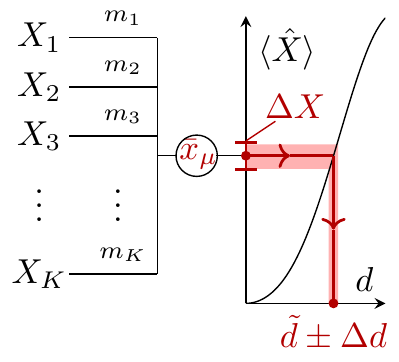}
\caption{
Moment-based estimation of the separation between the sources: the measured mean values of the available  observables are linearly combined with optimal coefficients and compared with a calibration curve.}
\label{Fig:scheme}
\end{figure}
The procedure discussed in this section to obtain the separation estimation $\tilde{d}$ from the optimal coefficients ${\bf m}$ in Eq.~\eqref{m} is illustrated in Fig.~\ref{Fig:scheme}. 

\section{Construction of the optimal observable for demultiplexing}
\label{Sec:construction}
Fisher-information based studies \cite{Tsang_PRX, Nair_2016} suggested that photon counting after spatial mode demultiplexing allows to approach the ultimate limit for the separation estimation. 
Therefore, in this Section we assume that we have access to $K$ spatial modes $\{v_k({\bf r})\}$ with associated field operators $\hat{a}_k$, and that we can measure the photon number in each mode $\hat{N}_k = \hat{a}_k^\dagger \hat{a}_k$. We use the formalism presented in Sec.~\ref{Sec:MethodOfMoments} to derive their optimal linear combination 
\begin{equation}
\hat{N}_{\bf m} = {\bf m}\cdot \hat{\bf N} = \sum_{i=1}^K m_i \hat{N}_i,
\end{equation}
with $\hat{\bf m}$ determined by equation \eqref{m}.
\added{Once this optimal observable is known, one only need to access its average value. The latter can be derived by measuring the mean photon number in each of the modes $\{v_k({\bf r})\}$ and then combining them according to the coefficients ${\bf m}$.}.

\subsection{Covariance matrix and derivative vector}
\label{Sec:Cov_and_dev}
\begin{figure}
\includegraphics{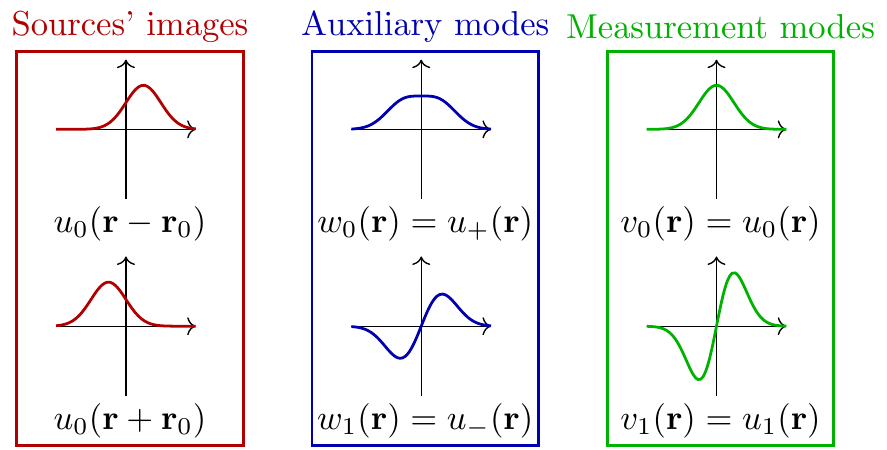}
\caption{The images of the two sources $u_0({\bf r} \pm {\bf r}_0)$ (red) are compared with the first two modes of the auxiliary mode basis $\{w_i ({\bf r}) \}$ (blue) and the ideal HG measurement basis $\{v_i ({\bf r}) = u_i ({\bf r})\}$ (green). The modes $\{v_i ({\bf r})\}$ are modified by misalignment and crosstalk.}
\label{Fig:modes}
\end{figure}

We now use the $P-$function~\eqref{P_image} to calculate the photon-number covariance matrix and derivative vector which are needed to compute the measurement sensitivity~\eqref{M} and coefficients~\eqref{m}.
To this goal, it is convenient to define an auxiliary mode basis $\{w_i({\bf r})\}$ obtained by extending $w_0({\bf r}) = u_+({\bf r})$ and $w_1({\bf r}) = u_-({\bf r})$ to a complete orthonormal basis.
Accordingly, for the field operators $\hat{b}_i$ associated with this basis, we have $\hat{b}_0 = \hat{b}_+$ and $\hat{b}_1 = \hat{b}_-$.
The field operators $\hat{a}_k$ in the measurement basis $\{v_k({\bf r})\}$ can be expanded in terms of the field operators $\hat{b}_k$ as $\hat{a}_k = \sum_l g_{kl} \hat{b}_l$, with $g_{kl} = \int d^2 {\bf r} v^*_{k}({\bf r})w_l({\bf r})$.
A one-dimensional comparison between the sources' images and the first two modes of the basis $\{w_k({\bf r})\}$ and $\{v_k({\bf r})\}$ is presented in Fig.~\ref{Fig:modes}.

Using the definition of the mean photon number in the measurement modes, we have
\begin{equation}
N_k = \langle \hat{N}_k\rangle = \langle \hat{a}^\dagger_k \hat{a}_k \rangle  =  \sum_{ij=\pm} g^*_{ki}g_{kj} \langle \hat{b}^\dagger_i \hat{b}_j \rangle,
\label{Nk}
\end{equation}
where we used that all $\hat{b}_{i\geq 2}$ are in the vacuum.
Analogously, the covariance matrix is given by $\Gamma_{kl}[d,\theta,\hat{\bf N}] = \langle \hat{N}_k \hat{N}_l \rangle - N_k N_l$ with
\begin{align}
\langle \hat{N}_k \hat{N}_l\rangle &=  \langle \hat{a}^\dagger_k \hat{a}^\dagger_l \hat{a}_k \hat{a}_l\rangle + \delta_{kl} \langle \hat{a}^\dagger_k \hat{a}_k \rangle \label{NN}\\ &= \sum_{mnpq=\pm}g_{mk}^*g_{nl}^*g_{pk}g_{ql} \langle\hat{b}^\dagger_m\hat{b}^\dagger_n\hat{b}_p\hat{b}_q \rangle +\delta_{kl}N_k. \nonumber
\end{align}
The normally ordered expectation values of the field operators $\hat{b}_i$ in Eqs.~\eqref{Nk} and \eqref{NN} can be calculated from the $P-$function \eqref{P_image} according to 
\begin{subequations}
\begin{align}
\langle \hat{b}^\dagger_m\hat{b}_n \rangle &= \int \beta_m^*\beta_n P(\beta_+,\beta_-) d^2 \beta_+ d^2 \beta_-,\\
\langle \hat{b}^\dagger_m \hat{b}^\dagger_n \hat{b}_p\hat{b}_q\rangle &= \int  \beta^*_m\beta^*_n\beta_p \beta_q P(\beta_+,\beta_-) d^2 \beta_+ d^2 \beta_-.
\end{align}
\label{integrals}
\end{subequations}
Given that $P(\beta_+,\beta_-)$ (see Eq.~\eqref{P_image}) is a Gaussian involving only the modes $u_\pm({\bf r})$, the only non-zero expectation values are
\begin{subequations}
\begin{align}
\langle \hat{b}^\dagger_\pm \hat{b}_\pm \rangle &= N_\pm,\\ 
\langle \hat{b}^\dagger_\pm \hat{b}_\mp \rangle &= -\gamma \sqrt{N_+N_-},\\
\langle \hat{b}^\dagger_\pm \hat{b}^\dagger_\pm \hat{b}_\pm \hat{b}_\pm \rangle &= 2 N_\pm^2,\\ 
\langle \hat{b}^\dagger_\pm  \hat{b}^\dagger_\mp \hat{b}_\pm \hat{b}_\mp \rangle &= N_+ N_-(1+\gamma^2),\\
\langle \hat{b}^\dagger_\pm \hat{b}^\dagger_\pm  \hat{b}_\mp \hat{b}_\mp \rangle &= 2\gamma^2 N_+ N_-,\\
\langle \hat{b}^\dagger_\pm \hat{b}^\dagger_\mp  \hat{b}_\mp \hat{b}_\mp \rangle &= -2\gamma N_\pm\sqrt{N_+ N_-}.
\end{align}
\label{bbb}
\end{subequations}
It is useful to write the mean photon number and the covariance matrices in terms of the sources images. 
For this purpose, we note that the coefficients $g_{k\pm}$ can be expressed in terms of the overlap functions with the source images $f_{\pm,k} = \int d{\bf r}v_k^*({\bf r})u({\bf r} \pm {\bf r}_0)$ as $g_{k\pm} = (f_{+,k} \pm f_{-,k})/\sqrt{2(1\pm\delta)}$, and using $N_{\pm} = N\kappa (1\pm \delta)$.
Accordingly, inserting Eqs.~\eqref{bbb} into Eqs.~\eqref{Nk} and \eqref{NN}, using $N_\pm = N\kappa (1\pm \delta)$, we get the mean photon numbers
\begin{equation}
N_k = N \kappa (|f_{+,k}|^2 + |f_{-,k}|^2) - \gamma N \kappa(|f_{+,k}|^2 - |f_{-,k}|^2)
\label{Nk_thermal}
\end{equation}
and the covariance matrix
\begin{equation}
\Gamma_{kl}[d,\theta,\hat{\bf N}] = \Gamma^0_{kl} + \gamma \Gamma^1_{kl} + \gamma^2  \Gamma^2_{kl},
\label{Gamma_thermal}
\end{equation}
with
\begin{subequations}
\begin{align}
\Gamma^0_{kl}&=(N\kappa)^2 (|f_{-,k}|^2|f_{-,l}|^2 + |f_{+,k}|^2|f_{+,l}|^2 \nonumber \\
&\quad+ 2 \Re [f_{-,k}f_{-,l}^*f_{+,l}f_{+,k}^*]) \label{Gamma0}\\
&\quad+\delta_{kl}  N \kappa (|f_{+,k}|^2 + |f_{-,k}|^2), \nonumber \\
\Gamma^1_{kl}&= - \delta_{kl}N \kappa(|f_{+,k}|^2 - |f_{-,k}|^2) ,\label{Gamma1}\\
\Gamma^2_{kl}&= (N\kappa)^2(|f_{-,k}|^2|f_{-,l}|^2 + |f_{+,k}|^2|f_{+,l}|^2 \nonumber\\
&\quad-2\Re[ f_{-,k}f_{-,l}^*f_{+,k}f_{+,l}^*]).  \label{Gamma2}
\end{align}
\end{subequations}
The derivative vector is obtained differentiating Eq.~\eqref{Nk_thermal} with respect to the parameter $d$, which gives
\begin{align}
D_k[d,\theta, \hat{\bf N}]  &= 2 N\kappa\left[ \Re \left(f_{+,k}^*\frac{\partial f_{+,k}}{\partial d} + f_{-,k}^*\frac{\partial f_{-,k}}{\partial d}\right)\right.\nonumber  \\
&\quad - \left.\gamma \Re \left(f_{+,k}^*\frac{\partial f_{+,k}}{\partial d} - f_{-,k}^*\frac{\partial f_{-,k}}{\partial d}\right)\right] \label{D_thermal}
\end{align}
In the case of two equally bright sources ($\gamma = 0$), we recover the expressions given in \cite{sorelli2021optimal}.
Analogously, setting $\gamma=0$ in Eq.~\eqref{Gamma_thermal}, we have $\Gamma_{kl}[d,\theta,\hat{\bf N}] = \Gamma^0_{kl}$ which coincides with the covariance matrix reported in \cite{sorelli2021optimal}.

\subsection{Noise sources}
\label{Sec:noise}
For the rest of this work, we focus on the case of a Gaussian PSF $u_0({\bf r}) = \sqrt{2/(\pi w^2)}\exp( - |{\bf r}|^2/w^2)$.
For this PSF, the quantum Cram\'er-Rao bound can be approached by demultiplexing Hermite-Gauss (HG) modes with width matching that of the PSF \cite{Tsang_PRX, Nair_2016}, i.e. $v_k({\bf r}) = u_k({\bf r}) \equiv u_{nm}({\bf r})$ with $k = (n,m)$,  such that $u_{00} ({\bf r}) = u_0 ({\bf r})$.
The HG modes are defined, for ${\bf r} = (x,y)$, as 
\begin{equation}
u_{nm}(x,y) = \mathcal{N}_{nm} H_n\left(\frac{\sqrt{2} x}{w}\right)H_m\left(\frac{ \sqrt{2} y}{w}\right)e^{-\frac{x^2+y^2}{w^2}},
\label{HG}
\end{equation}
where $H_n(x)$ are the Hermite polynomials, and $\mathcal{N}_{nm} = ((\pi/2)w^2 2^{n+m} n! m!)^{-1/2}$ is the normalization constant.
For ideal measurements in the HG mode basis, we have the overlap functions $f_{\pm,k} = \beta_{nm}(\pm {\bf r}_0)$, with 
\begin{equation}
\beta_{nm}({\bf a}) \equiv \int d^2 {\bf r} u_{nm}^*({\bf r}) u_{00}({\bf r} - {\bf a}),
\label{overlap}
\end{equation}
that fully determine the covariance matrix \eqref{Gamma_thermal} and the derivative vector\eqref{D_thermal}, and consequently the measurement sensitivity \eqref{M} and coefficients \eqref{m}.
A detailed discussion of the performances of our estimation strategy for ideal measurements is reported in Sec.~\ref{Sec:Moments_ideal}.

\begin{figure}
\includegraphics{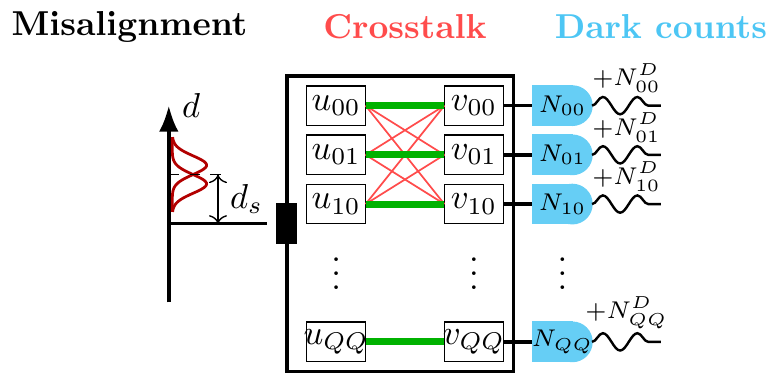}
\caption{Graphical illustration of a noisy demultiplexing measurement: the image of the two sources enters into a demultiplexer which performs mode sorting affected by crosstalk. Photon counting measurements affected by electronic noise are performed at each demultiplexer's output.}
\label{Fig:noise}
\end{figure}

We now study the impact of different noise sources on the covariance matrix \eqref{Gamma_thermal} and the derivative vector \eqref{D_thermal}.
In particular, we consider misalignment between the demultiplexing basis and the source centroid, as well as crosstalks between the detection modes.
These two imperfections affect Eqs.~\eqref{Gamma_thermal} and \eqref{D_thermal} only through modifications of the overlap funtions $f_{\pm,k}$.
Finally, we consider dark counts at the detection stage, that modify the diagonal of the covariance matrix~\eqref{Gamma_thermal}.
A schematic illustration of how the different noise sources enter a demultiplexing measurement is presented in Fig.~\ref{Fig:noise}.

\subsubsection{Misalignment}
The assumption that the demultiplexing basis is perfectly centered with respect to the centroid of the two sources is often not true in practice. 
In fact, the source centroid is in general a priori unknown and needs to be pre-estimated, possibly via direct imaging, to optimally align the demultiplexer.
In the case of faint sources, it was observed \cite{Tsang_PRX} that an imperfect positioning of the demultiplexer makes the Fisher information go to zero for small sources separations. 
Methods to mitigate this effect have been proposed by adaptively switching between demultiplexing and direct imaging \cite{Grace:20} or by optimizing the detection basis \cite{AlmeaidaPRA2021}. 
\added{An alternative approach, that we will not consider here, consists in estimating the source's centroid simultaneously with the separation \cite{Rehacheck2017}.} 

Within our model, a two-dimensional shift ${\bf r}_s = (d_s \cos\theta_s, d_s \sin\theta_s)^T$ of the centroid of the sources with respect to the demultiplexer axis enters the measurement sensitivity \eqref{M}, and the coefficients \eqref{m} through the overlap functions $f_{\pm,k} =\beta_{nm} (\pm{\bf r}_0 - {\bf r}_s)$.
As a trick, to compute the overlap functions $\beta_{nm} ({\bf a})$, we can use an analogy with the quantum mechanical harmonic oscillator.
In particular, we can interpret the integral~\eqref{overlap} as the overlap between the $(n,m)-$excited state of a two-dimensional harmonic oscillator and a coherent state (displaced to the phase-space coordinate ${\bf a}$) of the same harmonic oscillator. 
We therefore obtain
\begin{align}
&\beta_{nm} (\pm{\bf r}_0 + {\bf r}_s) = \nonumber\\
&= \frac{1}{\sqrt{n! m!}} e^{-\frac{\left(d_s \cos \theta_s \pm \frac{d}{2}\cos\theta\right)^2+\left(d_s \sin \theta_s \pm \frac{d}{2}\sin\theta\right)^2}{2w^2}}\nonumber\\
&\left(\frac{d_s}{w}\cos \theta_s\pm\frac{d}{2w}\cos \theta\right)^n\left(\frac{d_s}{w}\sin \theta_s\pm\frac{d}{2w}\sin \theta\right)^m.
\label{beta}
\end{align}

\subsubsection{Crosstalk} 
A recent experiment \cite{Boucher:20} identified imperfections in the mode decompositions as an important limitation for resolving the distance between incoherent point sources.
For sources with low brightness, it was reported that in the presence of crosstalk in the demultiplexing basis the Fisher information drops to zero for small separations \cite{gessner2020}.

To include the impact of crosstalk between the detection modes in our model, we follow \cite{gessner2020}, and we describe it as a unitary matrix $c_{kl}$ that maps the ideal HG modes $u_l({\bf r})$ into the actual measurement basis $v_k = \sum_l c_{kl} u_l({\bf r})$. 
Accordingly, the overlap functions $f_{\pm,k}$ are given by linear combinations of the functions \eqref{beta},
\begin{equation}
f_{\pm,mn} = \sum_{pq} c_{nm,pq} \beta_{pq} (\pm{\bf r}_0 - {\bf r}_s).
\label{f}
\end{equation}

To assess the impact of crosstalk on the sensitivity of our method, as well as on the shape of our optimal observable, we numerically generate $K \times K$ unitary matrices
\begin{equation}
C(\epsilon) = \exp \left(-i\epsilon \sum_{i=1}^{K^2-1} \lambda_i G_i \right),
\label{C}
\end{equation}
by sampling the uniformly distributed random real coefficients $\lambda_i$ ($0 \leq \lambda_i \leq 1$ and $\sum_i \lambda_i^2 = 1$), that multiply the generalized Gell-Mann matrices $G_i$ \cite{Bertlmann_2008}.
The positive parameter $\epsilon$ allows us to tune the crosstalk strength.
In experimentally relevant scenarios, crosstalk is generally weak \cite{Boucher:20,gessner2020}. 
Namely, the diagonal elements of the crosstalk matrix $C_{ii}(\epsilon) = c_{ii}$ are larger when compared to the off-diagonal ones $C_{i\neq j}(\epsilon) = c_{i \neq j}$.
Accordingly, we select $\epsilon$ to ensure  
\begin{equation}
\overline{|c_{ij}|^2} = \frac{1}{K(K-1)}\sum_{k\neq l = 1}^K |c_{kl}|^2 \ll 1 .
\end{equation}

In the following, when discussing results in the presence of crosstalk, we will consider multiple random realizations of the matrix $c_{ij}$, and quantify the average crosstalk probability via the ensemble average $\langle \overline{|c_{ij}|^2} \rangle$.

\subsubsection{Dark counts}
Electronic noise at the detection stage introduces additional photon counts, the dark counts, which contain no information on the parameter value $d$. 
Consequently, the signal-to-noise ratio in each detection mode is reduced by dark counts.
The impact of this noise source on distance estimation via spatial-mode demultiplexing has been investigated in the recent literature \cite{Len2020,Lupo2020,oh2020}.
Despite the different approaches, all these works obtained the same qualitative result: dark counts cause the Fisher information to drop to zero for small source separations.

To include dark counts in our model, we add to 
the quantum mechanical photon number operator in the measurement modes $\hat{N}_k$ a classical random variable $\xi_k$. 
In particular, $\xi_k$ are thermally distributed with mean value 
\begin{equation}
\langle \xi_k \rangle = N_k^{\rm dc}.
\end{equation}
To quantify the strength of dark counts with respect to the number of photons received in the image plane, we define $\sigma_k = N_k^{\rm dc}/2N\kappa$.
The measured mean photon number in each detection mode $N^\prime_k$ and the covariance matrix $\Gamma^\prime[d,\theta,\hat{\bf N}]$ are now obtained not only by computing quantum mechanical expectation values, but also classical averages over the probability distribution of $\xi_k$. 
Following this procedure, we obtain 
\begin{equation}
N^\prime_k =  N_k+  N_k^{\rm dc}
\end{equation}
and, given that $N_k^{\rm dc}$ is independent on $d$, the derivative vector $D[d,\theta,\hat{\bf N}]$ is untouched by dark counts.
On the other hand, the covariance matrix becomes 
\begin{equation}
\Gamma_{kl}^\prime[d,\theta,\hat{\bf N}] = \Gamma_{kl}[d,\theta,\hat{\bf N}] + \delta_{kl} N_k^{\rm dc}(N_k^{\rm dc} +1),
\end{equation}
with the extra diagonal term describing the noise induced by dark counts.

\section{Direct imaging}
\label{Sec:Direct_imaging}
In Secs. \ref{Sec:Moments_ideal} and~\ref{Sec:Additional_res}, we will discuss the performances of ideal and noisy demultiplexing, respectively, and we will compare them with those of direct imaging.
In order to perform such comparison, we now consider an ideal direct imaging system, and we evaluate its optimized moment-based sensitivity~\eqref{M} that bounds the Fisher information according to Eq.~\eqref{chain}. 

Direct imaging estimates the separation between the two sources from the intensity distribution in the image plane. 
The mean intensity is given by 
\begin{align}
I({\bf r}) &= \langle \hat{E}^{(+)\dagger} ({\bf r}){E}^{(+)} ({\bf r})\rangle \nonumber
\\&=  \sum_{i,j = \pm} u_i^*({\bf r}) u_j({\bf r}) \langle \hat{b}^\dagger_i \hat{b}_j \rangle 
\\&= N\kappa\left((1+\gamma)|u_0({\bf r}+{\bf r}_0)|^2 + (1-\gamma)|u_0({\bf r}-{\bf r}_0)|^2 \right). \nonumber
\end{align}
In the first step, we used the expansion of the electric field in the basis $\left \{ w_k({\bf r}) \right\}$ of the symmetric $ w_0({\bf r}) = u_+({\bf r})$ and antisymmetric modes $ w_1({\bf r}) = u_-({\bf r})$ (see Sec.~\ref{Sec:construction}), and the fact that,  in the quantum state \eqref{rhosSsA}, all higher order modes $w_{k \geq 2}({\bf r})$ are in the vacuum state.
In the second step, we used $\langle \hat{b}^\dagger_\pm \hat{b}_\pm \rangle = N_\pm$ and $\langle \hat{b}^\dagger_\pm \hat{b}_\mp \rangle = -\gamma \sqrt{N_+ N_-}$ (see Eqs.~\eqref{bbb}), as well as the relation \eqref{upm} between $u_0({\bf r} \pm {\bf r}_0)$ and $u_\pm({\bf r})$.

Let us now assume that the intensity measurements are performed with an ideal pixelized detector, with unity quantum efficiency and noiseless number-resolved photon counting at each pixel.
The pixels are defined by the area 
\begin{equation}
\mathcal{A}_{ij} = \{(x,y) :  |x -x_i| \leq x_p,  |y-y_i| \leq y_p  \},
\end{equation} 
with $2 x_p$ ($2 y_p$) the horizontal (vertical) size of the pixels and $(x_i, y_i)$ the coordinate of the center of the pixel.
We further assume that the total area $\mathcal{A}$ of the detector is fixed, and large enough to collect the full intensity in the image plane.
Accordingly, the pixel area is fully determined by the number of segments on each axis $N_p$, i.e. $x_p \times y_p =  \mathcal{A}/N_p^2$. 
The mean photon count per pixel is given by the mean intensity integrated over the pixel area
\begin{equation}
I_{ij} = N\kappa (1+\gamma) \Phi_{ij} + N\kappa (1-\gamma) \Psi_{ij},
\label{Iij}
\end{equation}
with 
\begin{subequations}
\begin{align}
\Phi_{ij} &= \int_{\mathcal{A}_{ij}} |u_0({\bf r} + {\bf r}_0)|^2 d^2 {\bf r},\\
\Psi_{ij} &= \int_{\mathcal{A}_{ij}} |u_0({\bf r} - {\bf r}_0)|^2 d^2 {\bf r}.
\end{align}
\label{PhiPsi}
\end{subequations}

The intensity coherence function is defined as
\begin{equation}
\Gamma({\bf r}, {\bf r}^\prime) =\langle \hat{E}^{(+)\dagger} ({\bf r}) {E}^{(+)} ({\bf r}) \hat{E}^{(+)\dagger} ({\bf r}^\prime){E}^{(+)} ({\bf r}^\prime) \rangle -  I({\bf r})I({\bf r}^\prime) 
\end{equation}
with
\begin{align}
&\langle \hat{E}^{(+)\dagger} ({\bf r})\hat{E}^{(+)} ({\bf r}) \hat{E}^{(+)\dagger} ({\bf r}^\prime) \hat{E}^{(+)} ({\bf r}^\prime) \rangle =\nonumber\\ &=\sum_{ijkl=\pm} u_i^*({\bf r}) u_j({\bf r}) u_k^*({\bf r}^\prime) u_l({\bf r}^\prime) \langle \hat{b}^\dagger_i \hat{b}_j \hat{b}^\dagger_k \hat{b}_l \rangle, \\
&=\sum_{ijkl=\pm} u_i^*({\bf r}) u_j({\bf r}) u_k^*({\bf r}^\prime) u_l({\bf r}^\prime) \langle \hat{b}^\dagger_i \hat{b}^\dagger_k \hat{b}_j \hat{b}_l \rangle +  \delta({\bf r} - {\bf r}^\prime)I ({\bf r})\nonumber
\end{align}
where in the last step we used the commutation relation $[\hat{b}_i, \hat{b}^\dagger_j] = \delta_{ij}$ and the completeness relation $\sum_{i} w_i^*({\bf r})w_i({\bf r}^\prime) = \delta({\bf r} - {\bf r}^\prime)$.
Using the expectation values~\eqref{bbb}, we can finally write the intensity coherence function as
\begin{equation}
\Gamma({\bf r}, {\bf r}^\prime) = \Gamma^0({\bf r}, {\bf r}^\prime) + \gamma\Gamma^1({\bf r}, {\bf r}^\prime)  + \gamma^2\Gamma^2({\bf r}, {\bf r}^\prime),
\label{Gamma_di}
\end{equation}
with
\begin{widetext}
\begin{subequations}
\begin{align}
 \Gamma^0({\bf r}, {\bf r}^\prime) &= N^2\kappa^2 \left(|u_0({\bf r} + {\bf r}_0)|^2 |u_0({\bf r}^\prime + {\bf r}_0)|^2 + |u_0({\bf r} - {\bf r}_0)|^2 |u_0({\bf r}^\prime - {\bf r}_0)|^2 \right.  \nonumber\\ 
 &\quad+ \left. 2\Re \{u^*_0({\bf r} + {\bf r}_0)u_0({\bf r} - {\bf r}_0)u^*_0({\bf r}^\prime - {\bf r}_0) u_0({\bf r}^\prime + {\bf r}_0) \} \right) \\
 &\quad+N\kappa\delta({\bf r} - {\bf r}^\prime)(|u_0({\bf r} + {\bf r}_0)|^2 + |u_0({\bf r} - {\bf r}_0)|^2 ), \nonumber\\
\Gamma^1({\bf r}, {\bf r}^\prime) &=N\kappa\delta({\bf r} - {\bf r}^\prime)(|u_0({\bf r} - {\bf r}_0)|^2 - |u_0({\bf r} - {\bf r}_0)|^2 ),\\
\Gamma^2({\bf r}, {\bf r}^\prime) &= N^2\kappa^2 \left(|u_0({\bf r} + {\bf r}_0)|^2 |u_0({\bf r}^\prime + {\bf r}_0)|^2 + |u_0({\bf r} - {\bf r}_0)|^2 |u_0({\bf r}^\prime - {\bf r}_0)|^2 \right.  \nonumber\\ 
&\quad- \left. 2\Re \{u^*_0({\bf r} + {\bf r}_0)u_0({\bf r} - {\bf r}_0)u^*_0({\bf r}^\prime - {\bf r}_0) u_0({\bf r}^\prime + {\bf r}_0) \} \right).
\end{align}
\end{subequations}
\end{widetext}

The covariance matrix for intensity measurements with the ideal pixelized detector described above is obtained by integrating the coherence function~\eqref{Gamma_di} over the pixel area according to
\begin{align}
\Gamma_{ijkl}[d,\theta,N_p]&= \int_{\mathcal{A}_{ij}} d^2 {\bf r} \int_{\mathcal{A}_{kl}} d^2 {\bf r}^\prime \Gamma({\bf r}, {\bf r}^\prime) \nonumber\\
&=\Gamma^0_{ijkl} + \gamma\Gamma^1_{ijkl}+\gamma^2\Gamma^2_{ijkl},
\label{Gamma_di_ijkl}
\end{align}
with
\begin{subequations}
\begin{align}
\Gamma^0_{ijkl} &= N^2\kappa^2 (\Phi_{ij} \Phi_{kl} + \Psi_{ij} \Psi_{kl} +2\Re\{\Xi^*_{ij}\Xi_{kl}\}) \nonumber \\ &+ \delta_{ik}\delta_{jl} N\kappa (\Phi_{ij} + \Psi_{ij}) \\
\Gamma^1_{ijkl} &= \delta_{ik}\delta_{jl} N\kappa (\Phi_{ij} - \Psi_{ij})\\
\Gamma^2_{ijkl} &= N^2\kappa^2 (\Phi_{ij} \Phi_{kl} + \Psi_{ij} \Psi_{kl} -2\Re\{\Xi^*_{ij}\Xi_{kl}\}),
\end{align}
\label{Gamma012_di}
\end{subequations}
where we have introduced 
\begin{equation}
\Xi_{ij} = \int_{\mathcal{A}_{ij}}  u^*_0({\bf r} + {\bf r}_0)u_0({\bf r} - {\bf r}_0) d^2 {\bf r}.
\label{Xi}
\end{equation}

The derivative vector is obtained from the mean intensity per pixel \eqref{Iij}, where the dependence on the parameter $d$ is only contained in the functions $\Phi_{ij}$ and $\Psi_{ij}$ \eqref{PhiPsi},
\begin{equation}
D_{ij}[d,\theta,N_p] = N\kappa (1+\gamma) \frac{\partial \Phi_{ij}}{\partial d} + N\kappa (1-\gamma)\frac{\partial \Psi_{ij}}{\partial d}.
\label{D_di}
\end{equation}
For a Gaussian PSF, the integrals \eqref{PhiPsi} and \eqref{Xi} can be evaluated analytically, and result in
\begin{subequations}
\begin{align}
\Phi_{ij} &= \frac{1}{4} \left(\text{erf}\left[\mathcal{C}_-(x_-)\right]-\text{erf}\left[\mathcal{C}_-(x^+_i)\right]\right)\\ \nonumber
&\quad\times\left(\text{erf}\left[\mathcal{S}_-(y^-_i)\right]-\text{erf}\left[\mathcal{S}_-(y^+_i)\right]\right),\\
\Psi_{ij} &= \frac{1}{4} \left(\text{erf}\left[\mathcal{C}_+(x^-_i)\right]-\text{erf}\left[\mathcal{C}_+(x^+_i)\right]\right)\\ \nonumber
&\quad\times\left(\text{erf}\left[\mathcal{S}_+(y^-_i)\right]-\text{erf}\left[\mathcal{S}_+(y^+_i)\right]\right),\\
\Xi_{ij} &= \frac{1}{4} e^{-\frac{d^2}{2 w^2}} \left(\text{erf}\left[x^-_i\right]-\text{erf}\left[ x^+_i\right]\right) \\ \nonumber
&\quad\times\left(\text{erf}\left[y^-_i\right]-\text{erf}\left[y^+_i\right]\right).
\end{align}
\label{GaussianPhiPsiXi}
\end{subequations}
In Eqs.~\eqref{GaussianPhiPsiXi}, we have defined 
\begin{subequations}
\begin{align}
\mathcal{C}_{\pm}(x) &= \frac{d \cos\theta}{\sqrt{2}w} \pm x,\\
\mathcal{S}_{\pm}(y) &= \frac{d \sin \theta}{\sqrt{2}w} \pm y,
\end{align}
\end{subequations}
and $x^{\pm}_i = \sqrt{2}(x_i \pm x_p)/w$, $y^{\pm}_i = \sqrt{2}(y_i \pm y_p)/w$.
Analogously, for the derivatives $\frac{\partial \Phi_{ij}}{\partial d}$ and $\frac{\partial \Psi_{ij}}{\partial d}$, we obtain 
\begin{subequations}
\begin{align}
\frac{\partial \Phi_{ij}}{\partial d}&=\frac{\cos \theta}{2 \sqrt{2 \pi } w} \left(e^{-\mathcal{C}^2_-(x^-_i)}-e^{-\mathcal{C}^2_-(x^+_i)}\right) \nonumber \\
&\quad \times \left(\text{erf}\left[S_-(y^-_i)\right]-\text{erf}\left[S_-(y^+_i)\right]\right)\nonumber\\&\quad
+\frac{\sin \theta}{2 \sqrt{2 \pi } w} \left(e^{-\mathcal{S}^2_-(y^-_i)}-e^{-\mathcal{S}^2_-(y^+_i)}\right) \nonumber \\
&\quad \times \left(\text{erf}\left[C_-(x^-_i)\right]-\text{erf}\left[C_-(x^+_i)\right]\right)\\
\frac{\partial \Phi_{ij}}{\partial d}&=\frac{\cos \theta}{2 \sqrt{2 \pi } w} \left(e^{-\mathcal{C}^2_+(x^-_i)}-e^{-\mathcal{C}^2_+(x^+_i)}\right) \nonumber \\
&\quad \times \left(\text{erf}\left[\mathcal{S}_+(y^-_i)\right]-\text{erf}\left[\mathcal{S}_+(y^+_i)\right]\right)\nonumber\\&\quad
+\frac{\sin \theta}{2 \sqrt{2 \pi } w} \left(e^{-\mathcal{S}^2_+(y^-_i)}-e^{-\mathcal{S}^2_+(y^+_i)}\right) \nonumber \\
&\quad \times \left(\text{erf}\left[\mathcal{C}_+(x^-_i)\right]-\text{erf}\left[\mathcal{C}_+(x^+_i)\right]\right).
\end{align}
\label{DerivativePhiPsi}
\end{subequations}
The covariance matrix \eqref{Gamma_di_ijkl} and the derivatives vector \eqref{D_di}, together with Eqs. \eqref{GaussianPhiPsiXi} and \eqref{DerivativePhiPsi}, generalize the results in the Supplementary Material of \cite{Nair_2016} to arbitrary two-dimensional arrangements and unequal brightnesses of the two sources.

\begin{figure}
\includegraphics[width=\columnwidth]{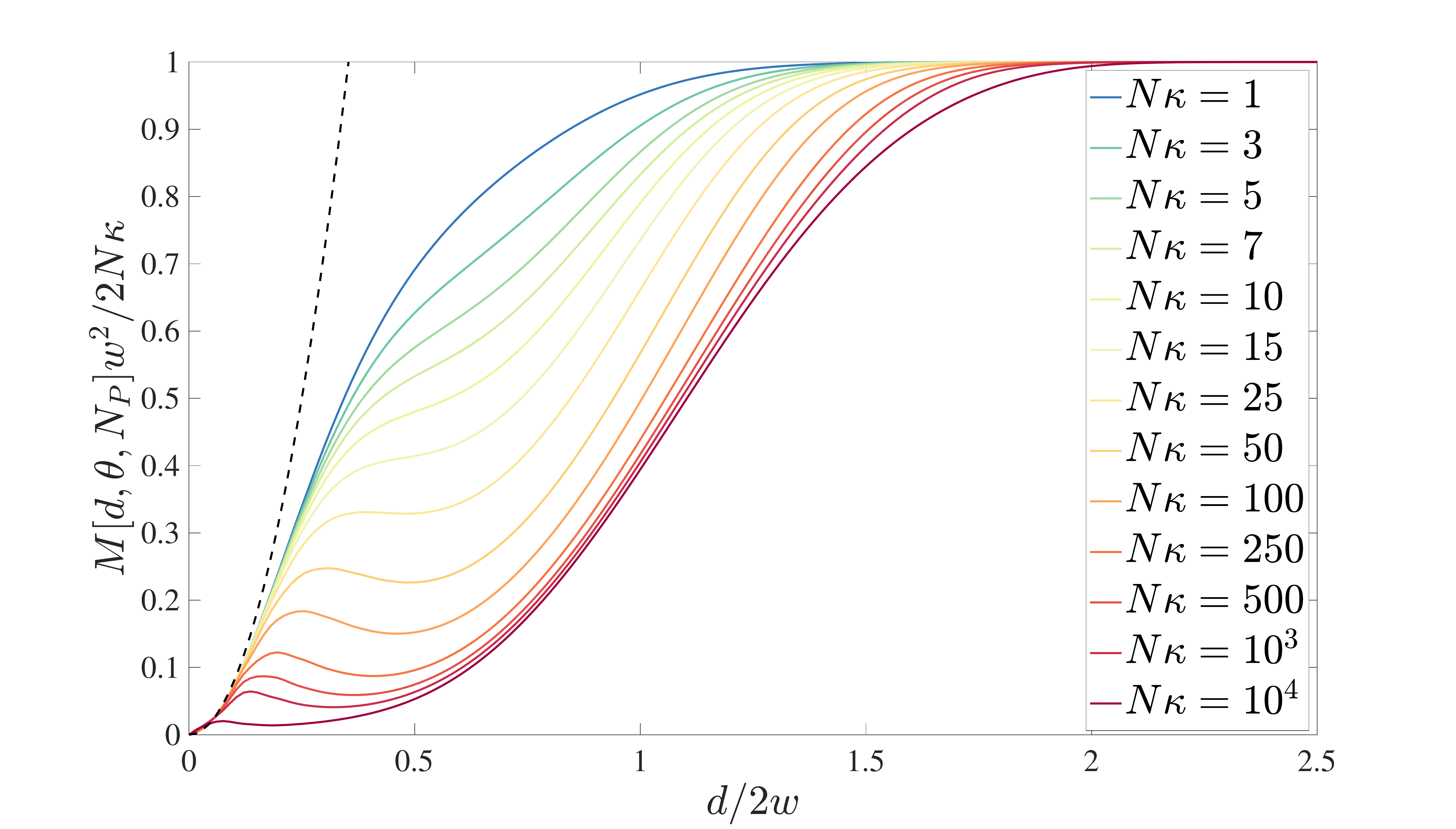}
\caption{Direct imaging sensitivity $M[d,\theta,N_p]$ for two equally bright thermal sources ($\gamma = 0$) with different mean number of received photons $N\kappa$. 
We considered an alignment angle $\theta = \pi/4$ and a square detector of side $6w$ divided into $N_p = 50\times 50$ pixels.
The black dashed line represent the analytical approximation for small source separation given by Eq.~\eqref{M_di_small}.}
\label{Fig:M_di}
\end{figure}
Let us now note that the covariance matrix~\eqref{Gamma_di_ijkl}, can be rewritten in the form 
\begin{equation}
\Gamma[d,\theta,N_P] = I +UU^T,
\end{equation}
where $I$ is a diagonal matrix whose elements are given by the mean intensities per pixels $I_{k=ij}$ as reported in Eq.~\eqref{Iij}, and $U = ({\bf \nu}^{(1)}, {\bf \nu}^{(2)},{\bf \nu}^{(3)})^T$ is $N_p^2\times 3$ matrix whose columns are given by the vectors 
\begin{subequations}
\begin{align}
\nu^{(1)}_{k=ij} &= N\kappa \sqrt{1+\gamma^2} \Phi_{ij},\\
\nu^{(2)}_{k=ij} &= N\kappa \sqrt{1+\gamma^2} \Psi_{ij},\\
\nu^{(3)}_{k=ij} &= N\kappa \sqrt{2(1-\gamma^2)} \Xi_{ij},
\end{align}
\end{subequations}
where we used that according to Eqs.~\eqref{GaussianPhiPsiXi} the functions $\Phi_{ij}$, $\Psi_{ij}$ and $\Xi_{ij}$ are all real.
Then, using the Woodbury matrix identity \cite{hager1989updating}, we can write the inverse of the covariance matrix as 
\begin{equation}
\Gamma^{-1}_{ijkl}[d,\theta,N_P] = \frac{\delta_{ik}\delta_{jl}}{I_{ij}} -
\Lambda_{ijkl},
\label{Gamma_di_inv}
\end{equation}
where the correction $\Lambda = U(\mathbbm{1}_3 + U^TU)^{-1}U^T$, with $\mathbbm{1}_3$ the $3\times 3$ matrix identity, can be evaluated numerically.
In our numerical calculations, independently of the mean photon number of the two thermal sources, we did not observe appreciable changes in  the behaviour of the measurement sensitivity $M[d,\theta,N_p]$ for $N_p \gtrsim 50\times50$.
Accordingly, $N_p = 50\times50$ was used to obtain all direct imaging curves reported in this work.

The representation~\eqref{Gamma_di_inv} allows to write the optimal direct imaging sensitivity as 
\begin{align}
M[d,\theta,N_p] &= \sum_{ij=1}^{N_p} \frac{1}{I_{ij}}\left( \frac{\partial I_{ij}}{\partial d}\right)^2 \label{M_di}\\
&\quad -\sum_{ijkl=1}^{N_p} D_{ij}[d,\theta,N_p] \Lambda_{ijkl}D_{kl}[d,\theta,N_p], \nonumber
\end{align}
where the first term is of order $N\kappa$, while the second one is of order $(N\kappa)^2$ and negative. 
As a consequence, the latter tends to reduce the relative sensitivity when increasing the source brightness.
This behaviour can be observed in Fig.~\ref{Fig:M_di}, where the direct imaging sensitivity $M[d,\theta,N_p]$ is plotted for different mean photon numbers of the sources.

The limit of continuous direct imaging is obtained for $N_p \to \infty$. 
In this limit, the first term in Eq.~\ref{M_di} becomes 
\begin{equation}
\mathcal{F}_{DI}[d,\theta] = \int  \frac{1}{I({\bf r})}\left(\frac{\partial I({\bf r})}{\partial d} \right)^2 d^2 {\bf r},
\label{F_di}
\end{equation}
which coincides with the direct imaging Fisher information calculated assuming Poissonian sources in the $N\kappa \ll 1$ regime \cite{Tsang_PRX}.
Accordingly, when the number of photons in the image plane is low ($N\kappa \ll 1$), when the second term in Eq.~\eqref{M_di} is negligible, the optimized moment-based sensitivity saturates the Cram\'er-Rao bound for continuous direct imaging.

In Fig.~\ref{Fig:M_di}, we see that the reduction of the measurement sensitivity due the $\Lambda$ term in Eq.~\eqref{M_di} is relevant only for intermediate source separations.
In fact, the behaviour of the measurement sensitivity $M[d,\theta,N_p]$~\eqref{M_di} for small distances between the sources is dominated by the diagonal part of the covariance matrix~\eqref{Gamma_di_ijkl}.
Accordingly, in this regime, we can approximate the direct imaging sensitivity by considering an expansion of the integral~\eqref{F_di} for $d/2w \ll 1$. 
Such an expansion can be performed analytically and results in 
\begin{equation}
M_{DI}[d,\theta] = \frac{2N\kappa}{w^2}\left(\gamma^2 + 4x^2(2 - 5\gamma^2 + 3\gamma^4)\right)+ O(x^4).
\label{M_di_small}
\end{equation}
The approximation~\eqref{M_di_small} is compared (for $\gamma =0$) with the exact numerical results for different mean photon numbers in Fig.~\ref{Fig:M_di}.

\begin{figure}
\includegraphics[width = \columnwidth]{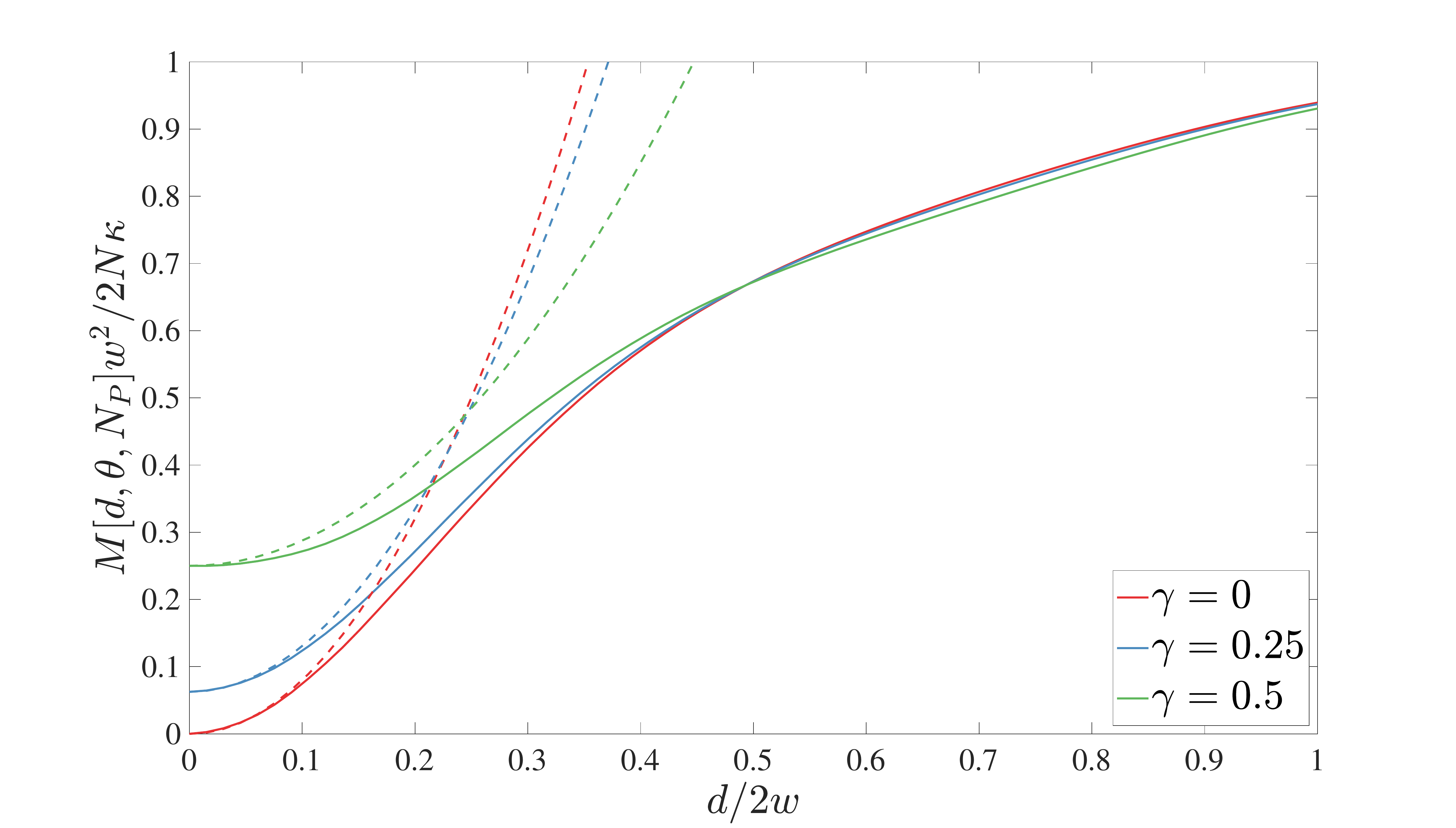}
\caption{Direct imaging sensitivity $M[d,\theta,N_p]$ for thermal sources with fixed total mean photon number $N\kappa = 1.5$ and different brightness imbalances: $\gamma =0$ (red), $\gamma =0.25$ (blue), $\gamma =0.5$ (green). 
We considered an alignment angle $\theta = \pi/4$ and a square detector of length $6w$ divided into $N_p = 50 \times 50$ pixels.
The dashed lines represent the analytical approximation for small source separation given by Eq.~\eqref{M_di_small}.}
\label{Fig:DI_gamma}
\end{figure}
It is interesting to note that for $\gamma \neq 0$, Eq.~\eqref{M_di_small} does not vanish for $d =0$ (see Fig.~\ref{Fig:DI_gamma}).
Accordingly, an asymmetry in the intensity distribution between the two sources eases the separation estimation.
In particular, we see that for extreme intensity imbalances $\gamma \to 1$, the direct imaging sensitivity for small separations tends to its maximal value $M_{DI}[d,\theta] \to 2N\kappa/w^2$.
We can understand this result considering that in the $\gamma \to 1$ limit the estimation of the sources separation becomes the localization of a single source.

\section{Ideal demultiplexing}
\label{Sec:Moments_ideal}
In this Section, we present analytical expressions for the optimal moment-based sensitivity, according to Eq.~\eqref{M}, attainable with ideal demultiplexing measurements and for the optimal observable achieving it, as given by Eq.~\eqref{m}.

\subsection{Analytical inversion of the covariance matrix}
\label{Sec:inversion}
We now focus on the case of ideal intensity measurements of $K= (Q+1)^2$ HG modes $u_{nm}({\bf r})$ with $0 \leq n,m \leq Q$.
In this case, the overlap functions $\beta_{nm} (\pm{\bf r}_0)$ are obtained setting $x_s = y_s = 0$ in Eq.~\eqref{beta}, which implies the symmetry
\begin{equation}
\beta_{nm} (-{\bf r}_0) = (-1)^{n+m} \beta_{nm} ({\bf r}_0).
\label{symmetry}
\end{equation} 
Equation~\eqref{symmetry} causes the mean photon number in the measurement modes $N_k$ (see Eq.~\eqref{Nk_thermal}) to be independent of the photon-number imbalance $\gamma$:
\begin{equation}
N_{nm} = 2 N\kappa \beta^2_{nm}({\bf r}_0).
\label{N_ideal}
\end{equation}
Substituting Eq.~\eqref{symmetry} into Eq.~\eqref{Gamma_thermal} also simplifies the covariance matrix, which becomes
\begin{align}
 \Gamma = {\rm diag}({\bm \nu}) + {\bm \xi}{\bm \xi}^T/2 + {\bm \zeta}{\bm \zeta}^T/2,
 \label{Gamma_ideal}
\end{align}
with $ {\rm diag}({\bm \nu})$ a diagonal $K\times K$ matrix whose elements are given by the vector ${\bm \nu}_{k=(n,m)} = N_{nm}$ (see Eq.~\eqref{N_ideal}), while the other two terms are outer products of vectors with elements ${\bm \xi}_{k=(n,m)} = \sqrt{1+\gamma^2}N_{nm}$ and ${\bm \zeta}_{k=(n,m)} =(-1)^{n+m} \sqrt{1-\gamma^2}N_{nm}$.
This particular form of the covariance matrix allows for its analytical inversion by two successive applications of the Sherman-Morrison formula  \cite{hager1989updating}. 
Accordingly, we obtain
\begin{align}
&\Gamma^{-1}_{mnm^\prime n^\prime} = \frac{\delta_{m m^\prime} \delta_{n n^\prime}}{2N\kappa\beta_{mn}^2({\bf r}_0)} \label{GammaInv}\\
&- \frac{(-1)^{m+n+m^\prime+n^\prime}A_+ - B((-1)^{m+n} + (-1)^{m^\prime +n^\prime}) + A_-}{A_+A_--B^2}, \nonumber
\end{align}
with 
\begin{subequations}
\begin{align}
A_{\pm} &= \frac{2}{1\pm\gamma^2} + 2N\kappa\sum_{mn=0}^Q\beta_{mn}^2({\bf r}_0), \label{Apm}\\
B &= 2N\kappa \sum_{mn=0}^Q(-1)^{m+n}\beta_{mn}^2({\bf r}_0) \label{B}.
\end{align}
\end{subequations}

Equation~\eqref{GammaInv} is valid only under the assumption that the covariance matrix is invertible. 
According to the Sherman-Morrison formula, this is the case if and only if ${\rm diag}({\bm \nu})$ is invertible, i.e. when $\beta_{nm}({\bf r}_0) \neq 0$ for every $n,m$.
Looking at Eq.~\eqref{beta}, for $x_s = y_s =0$, we see that this condition is not satisfied for $d=0$ or $\theta =k\pi/2$, with $k= 0, 1, 2, \dots$.
Let us first focus on $d=0$.
In this case, $\beta_{nm}({\bf r}_0) =0$ for all $n,m$. However, this case corresponds to having only one source and the estimation problem under study is not well posed.
On the other hand, for $\theta =k\pi/2$ with $k$ even (odd), we have $\beta_{nm}({\bf r}_0) =0\; \forall n(m)\neq 0$. 
We can therefore make the covariance matrix \eqref{Gamma_ideal} invertible by removing all modes with $n(m) =0$ from our measurement basis, and carry out the estimation within this smaller set of modes.
This fact has a clear physical interpretation: 
The HG modes (see Eq.~\eqref{HG}) have the form $u_{nm}(x,y) = h_n(x) h_m(y)$, accordingly when $\theta$ is an odd (even) multiple of $\pi/2$ the two sources are aligned with the $x-$($y-$)axis and the only relevant modes are $u_{n0}(x,y) = h_n(x)$ ($u_{0m}(x,y) = h_m(y)$).
Equation \eqref{GammaInv} yields the correct results also for $\theta =k\pi/2$ by extracting only the relevant matrix elements.

\subsection{Measurement sensitivity}
The measurement sensitivity $M[d, \theta, \hat{\bf N}]$ ~\eqref{M}, which is the maximal sensitivity achievable by any linear combination of the mean photon numbers in the measurement modes $N_k$ \cite{GessnerPRL2019}, can be obtained  by combining the inverse covariance matrix~\eqref{GammaInv} with the derivative vector
\begin{equation}
D_{mn} [d,\theta,\hat{\bf N}]= \frac{2\kappa N}{w x} (m+n-x^2)\beta^2_{mn}({\bf r}_0),
\label{D_ideal}
\end{equation}
with $x = d/2w$, and is given by
\begin{equation}
M[d,\theta,\hat{\bf N}] = \frac{2 N \kappa}{w^2} \left[F -(2 N\kappa)(\delta_1 +\delta_2 + \delta_3) \right],
\label{M_ideal}
\end{equation}
with 
\begin{subequations}
\begin{align}
F &= \sum_{mn=0}^Q\frac{(n + m -x^2)^2}{x^2}\beta_{mn}^2({\bf r}_0), \label{F}\\
\delta_1 &= \frac{A_+}{A_+A_- - B^2}\mathcal{S}_1^2, \label{delta1}\\
\delta_2 &=-\frac{2B}{A_+A_- - B^2}\mathcal{S}_1\mathcal{S}_2,\label{delta2}\\
\delta_3 &=\frac{A_-}{A_+A_- - B^2}\mathcal{S}_2^2. \label{delta3}
\end{align}
\end{subequations}
and 
\begin{subequations}
\begin{align}
\mathcal{S}_1 &= \sum_{p,q=0}^Q (-1)^{p+q}\frac{p+q -x^2}{x}\beta^2_{pq}({\bf r}_0),\\
\mathcal{S}_2&= \sum_{p,q=0}^Q \frac{p+q -x^2}{x}\beta^2_{pq}({\bf r}_0).
\end{align}
\label{S+S-}
\end{subequations}

We immediately see that, when the number of received photons is low ($N\kappa \ll 1$), the sensitivity $M[d,\theta,\hat{\bf N}]$ is dominated by the $F$ term. 
This term coincides with the Fisher information calculated assuming Poissonian sources in the $N\kappa \ll 1$ regime (compare Eq.~\eqref{F} with Eq. (21) in the Supplementary Material of \cite{gessner2020}).
Moreover, $F$ does not depend on $\gamma$, i.e. the number of received photons is low, the sensitivity in the estimation of $d$ does not depend on the relative brightness of the two sources.

For higher received photon numbers, the terms of order $(N\kappa)^2$ become relevant.
Given that the functions $\delta_1, \delta_2$ and $\delta_3$ are positive, the quadratic terms in the photon number always reduce the measurement sensitivity.
Moreover, they depend on $\gamma$ through $A_\pm$. 
Accordingly, the loss of sensitivity appearing for higher received photon numbers depends on how the photons are distributed between the two sources.

Let us now consider the limiting case where the full HG basis is measured, i.e. $Q \to \infty$.
In this limit, $F \to 1$, $\delta_2$ and $\delta_3$ vanish, while $\delta_1$ remains finite and provides a correction to the measurement sensitivity, which takes the form

\begin{align}
&M_{\rm inf}[d,\theta,\hat{\bf N}] = \lim_{Q \to  \infty} M[d,\theta, \hat{\bf N}] =  \label{M_inf}\\
&\frac{2N\kappa}{w^2}\left[1 - \frac{4 \left(1-\gamma ^2\right) N\kappa  e^{-4 x^2} x^2 \left(\left(\gamma ^2+1\right) N\kappa +1\right)}{\left(1-\gamma ^4\right) N^2\kappa ^2 \left(1-e^{-4 x^2}\right)+2 N\kappa+1}\right].\nonumber
\end{align}

\begin{figure}[t]
\includegraphics[width=\columnwidth]{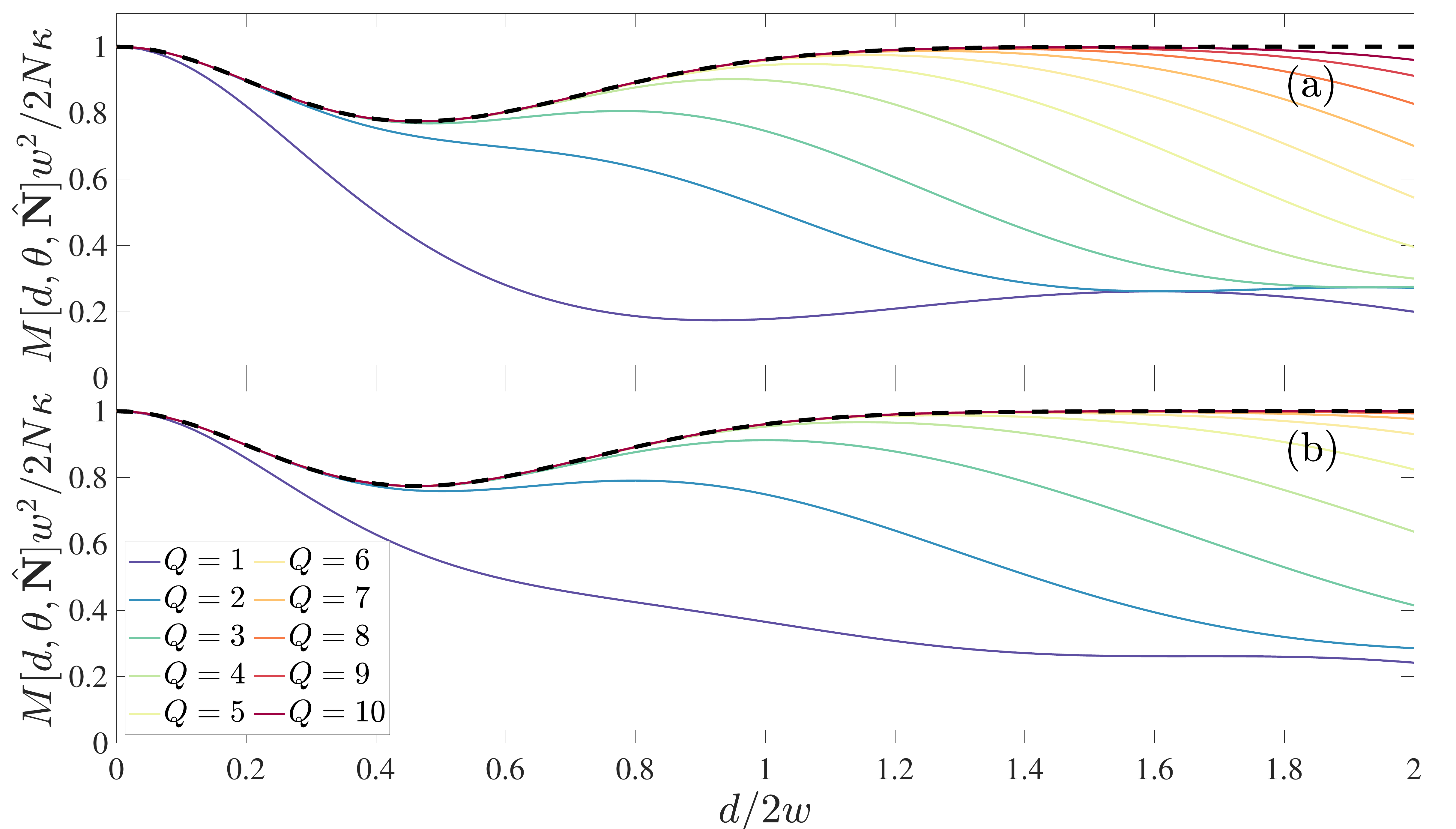}
\caption{Measurement sensitivity Eq.~\eqref{M_ideal} for equally bright sources ($\gamma = 0$) and for different values of $Q$ (solid lines) and alignment angle $\theta = 0$ (a) and $\theta = \pi/4$(b). The black dashed line represents the quantum Fisher information  Eq.~\eqref{Minf0}. We assumed $N\kappa = 1.5$ for both panels.}
\label{FIg:M_Q}
\end{figure}
\begin{figure}[t!]
\includegraphics[width=\columnwidth]{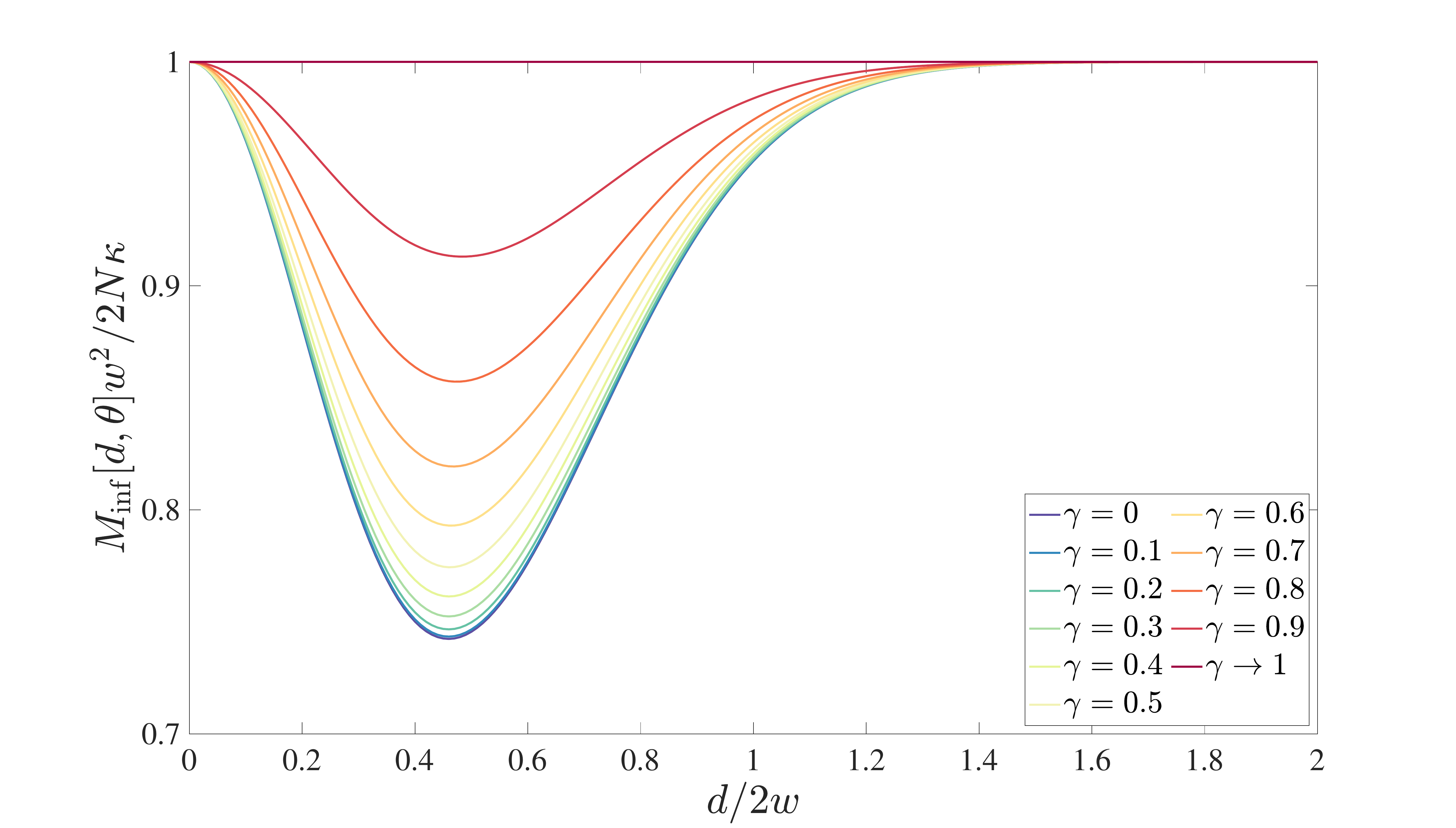}
\caption{Asymptotic measurement sensitivity Eq.~\eqref{M_inf} for fixed $N\kappa = 1.5$ and different brightness imbalances $\gamma$.}
\label{Fig:M_gamma}
\end{figure}

The equal brightness case is obtained by setting $\gamma = 0$ in Eq.~\eqref{M_inf}, which results in
\begin{equation}
M_{\rm inf}[d,\theta]  =  \frac{2N\kappa}{w^2} - \frac{8 N^2 \kappa ^2 x^2 (N\kappa  +1)e^{-4 x^2}}{w^2\left((N \kappa +1)^2-N^2\kappa ^2e^{-4 x^2}\right)},
\label{Minf0}
\end{equation}
which coincides with the quantum Fisher information for two equally bright thermal sources calculated in \cite{Nair_2016,LupoPirandola}.
Accordingly, our estimation strategy saturates the quantum Cram\'er-Rao bound in the asymptotic limit.
The second term in Eqs.~\eqref{M_inf} and \eqref{Minf0} reduces the estimation sensitivity for intermediate separations, i.e. a dip appears in Figs. \ref{FIg:M_Q} and \ref{Fig:M_gamma}. 
This dip gets deeper when increasing the mean photon number $N\kappa$ (see also \cite{Nair_2016, LupoPirandola}).
\added{Physically, this is due to the fact, that for higher brightness of the sources there is a non negligible probability of detecting multiple photons in the same modes which reduces the information that can be extracted from each photon.}

The behaviour of the measurement sensitivity \eqref{M_ideal} for $\gamma = 0$ and different values of $Q$ is compared with the quantum Fisher information \eqref{Minf0} in Fig.~\ref{FIg:M_Q}. 
While the asymptotic expression \eqref{Minf0} is independent on the alignment angle $\theta$,  this is not true at finite $Q$. In particular, $M[d,\theta, \hat{\bf N}]$ is minimal when the sources are aligned with the axes ($\theta = 0, \pi/2$), and maximal when the sources are aligned along a bisector ($\theta = \pi/4, 3 \pi/4)$.
This behaviour can be observed in Fig.~\ref{FIg:M_Q}, where in panel (a) ($\theta = 0$), for large separations $d$, the measurement sensitivity is clearly distinguishable from the quantum Fisher information for all considered values of $Q$. 
On the other hand, in panel (b) ($\theta = \pi/4$), the measurement sensitivity cannot be discerned from the quantum Fisher information for $Q \geq 9$.
A similar behaviour was reported for the Fisher information, for the regime of low photon numbers in the image plane, in the Supplementary Material of \cite{gessner2020}.
The optimality of $\theta = \pi/4$ is due to the fact that we assumed that we have a limited and equal number of modes available in each direction. 
Under this assumption, aligning the sources in one direction, we can use only $Q$ of the $Q^2$ available modes. 
This limits the resolution especially for large separations. 
On the other hand, aligning the sources along the bisector, is equivalent to rotating the basis in order to have all $Q^2$ available modes along the sources axis, which provides the optimal resolution.

The asymptotic behaviour ($Q\to\infty$) of the measurement sensitivity~\eqref{M_inf} for different brightnesses of the two sources ($\gamma \neq 0$) is illustrated in Fig.~\ref{Fig:M_gamma}.
We can see that the dip at intermediate separations gets shallower when increasing $\gamma$, and it disappears in the limit $\gamma \to 1$.
Accordingly, as observed for direct imaging in Sec.~\ref{Sec:Direct_imaging}, an intensity unbalance between the sources makes it easier to estimate their separation.

\subsection{Measurements coefficients}
Combining Eqs.~\eqref{GammaInv} and \eqref{D_ideal} according to Eq. \eqref{m}, we obtain the expression for the measurement coefficients (dashed lines in Fig.~\ref{Fig:coeff} in Sec.~\ref{Sec:Additional_res})
\begin{align}
\frac{w}{\eta}m_{ij} &= \frac{i+j -x^2}{x} - \frac{2\kappa N}{A_+A_- - B^2}\times \label{m_th} \\&\quad \left[ ((-1)^{i+j}A_+ - B)\mathcal{S}_+ -  ((-1)^{i+j}B- A_-)\mathcal{S}_-\right] \nonumber
\end{align}
with the normalization constant 
\begin{equation}
\eta[d,\theta,Q] = \left(\sum_{i,j = 0}^Q m_{ij}^2\right)^{-1/2}.
\label{eta}
\end{equation}
From Eq.~\eqref{m_th}, we note that the coefficients of the optimal observable only depend on the sum of the indices $i + j$.
Accordingly, independently on the sources orientations or their relative intensities, the method of moments prescribes to measure with the same weight the photon number in all HG modes of the same order.
In other words, in the absence of noise, it is optimal to measure a circularly symmetric intensity distribution, which achieves the same sensitivity for each alignment angle $\theta$.

The above observation does not apply when the sources are aligned along the $x-$($y-$)axis. 
In these cases, as discussed in Sec. \ref{Sec:inversion}, one needs only to measure modes with $i(j)=0$.
The correct coefficients can be still obtained from Eq.~\eqref{m_th} by setting $i= 0$($j =0$). 
Further comments on the behaviour of the measurement coefficients \eqref{m_th} is postponed to Sec.~\ref{Sec:Additional_res}, where we will compare them with their counterpart in presence of noise. 

\section{Noisy demultiplexing}
\label{Sec:Additional_res}

\subsection{Measurement sensitivity}
\label{Sec:noisy_sensitivity}
Let us start our discussion on the sensitivity of our estimation strategy in the presence of noise by commenting on its optimality.
We have already observed in Sec.~\ref{Sec:Moments_ideal} that for ideal demultiplexing our method approaches the quantum Cram\'er-Rao bound for arbitrary source separations and brightnesses, when a sufficiently large number of modes is measured.
Moreover, when the number of received photons is low ($N\kappa \ll 1$), the covariance matrix \eqref{Gamma_thermal} is dominated by its diagonal terms, and we have
\begin{equation}
M[d,\theta, \hat{\bf N}] \approx \sum_{k=1} \frac{1}{N_k} \left(\frac{\partial N_k}{\partial d} \right)^2.
\label{M_small}
\end{equation}
As discussed in Sec.~\ref{Sec:noise}, crosstalk and misalignment only affect the overlap functions $f_{\pm,k}$, while dark counts modify the diagonal of the covariance matrix. 
As a consequence, when all noise sources are considered, Eq.~\eqref{M_small} remains valid if we substitute the mean photon number $N_k$ with the mean photon number plus noise: $N_k^\prime = N_k + N_k^{\rm dc}$. 
Equation~\eqref{M_small} is equal to the Fisher information for demultiplexing calculated assuming Poissonian sources in the low brightness regime ($N\kappa \ll 1$) \cite{Tsang_PRX}.
Accordingly, for $N\kappa \ll 1$ our estimation strategy saturates the Cram\'er-Rao bound even in the presence of noise.

\begin{figure}[t]
\includegraphics[width=\columnwidth]{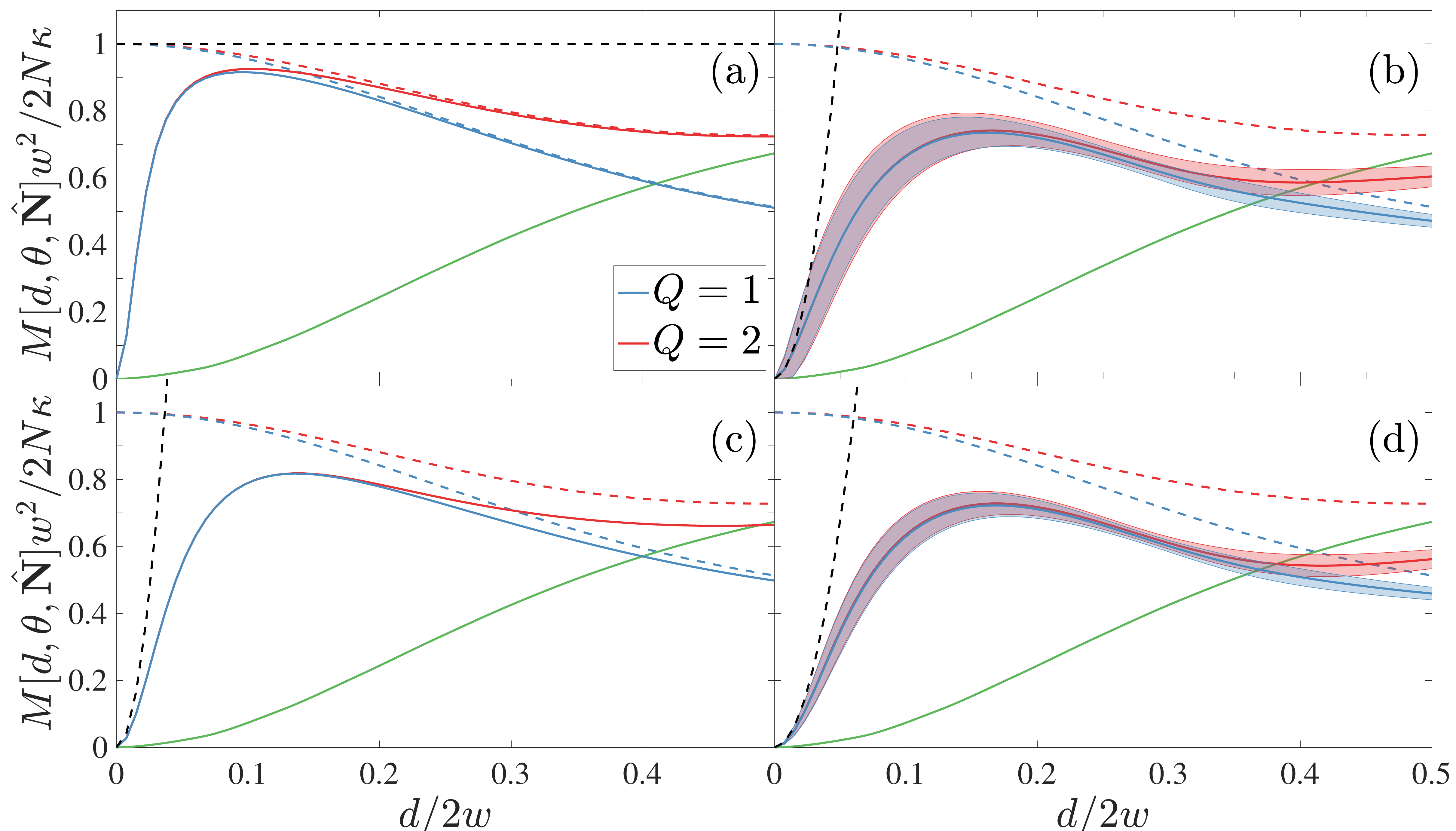}
\caption{
Measurement sensitivity $M[d,\theta,\hat{\bf N}]$ for demultiplexing into HG modes $u_{nm}({\bf r})$ with $n,m \leq Q$ ($K = (Q+1)^2$) with $Q=1$ (red solid line) and $Q=2$ (blue solid line) with different noise sources: (a) misalignment ($d_s/2w = 0.01, \theta_s = \pi/4$), (b) crosstalk ($\langle \overline{|c_{ij}|^2} \rangle = 0.0017$), (c) dark counts ($\sigma_k = 0.001\; \forall k$), and (d) all three noise sources at the same time.
Black dashed lines represent short-distances approximations, in particular, we used Eq.~\eqref{M_xs} in (a), Eq.~\eqref{M_ct} with $|t|^2 =\langle \overline{|c_{ij}|^2} \rangle$, and $\sigma=0, 0.001$ in (b) and (d) respectively, and Eq.~\eqref{M_dc} with $\sigma=0.001$ in (c).
Red and blue dashed lines show the results for ideal measurements \eqref{M_ideal}, for $Q=1$ and $2$ respectively.
The green solid line describes direct imaging results \eqref{M_di}.
For all plots, we assumed $N\kappa = 1.5$, $\theta = \pi/4$ and $\gamma =0$.}
\label{Fig:M_zoom}
\end{figure}
For thermal sources of arbitrary brightnesses and demultiplexing (noisy or not) in a finite number $K$ of spatial modes, the Fisher information is unknown. 
In this regime, the sensitivity of our method provides a lower bound for the Fisher information.
The latter can be saturated in practice by applying the simple estimation strategy presented in Sec.~\ref{Sec:MethodOfMoments} to the optimal linear combination of mean photon numbers in the measurement modes discussed in Sec.~\ref{Sec:observable}.
The measurement sensitivity $M[d,\theta,\hat{\bf N}]$ is presented in Fig.~\ref{Fig:M_zoom} for different numbers of measured modes and different noise levels.
We can see that all noise sources reduce the sensitivity with respect to its ideal value~\eqref{M_ideal} (red and blue dashed curves in Fig.~\ref{Fig:M_zoom}). 
In particular, we have that for equally bright sources, $M[d,\theta,\hat{\bf N}]$ vanishes for $d \to 0$, and accordingly it gets harder to resolve small source separations.
\begin{figure}[t!]
\includegraphics[width=\columnwidth]{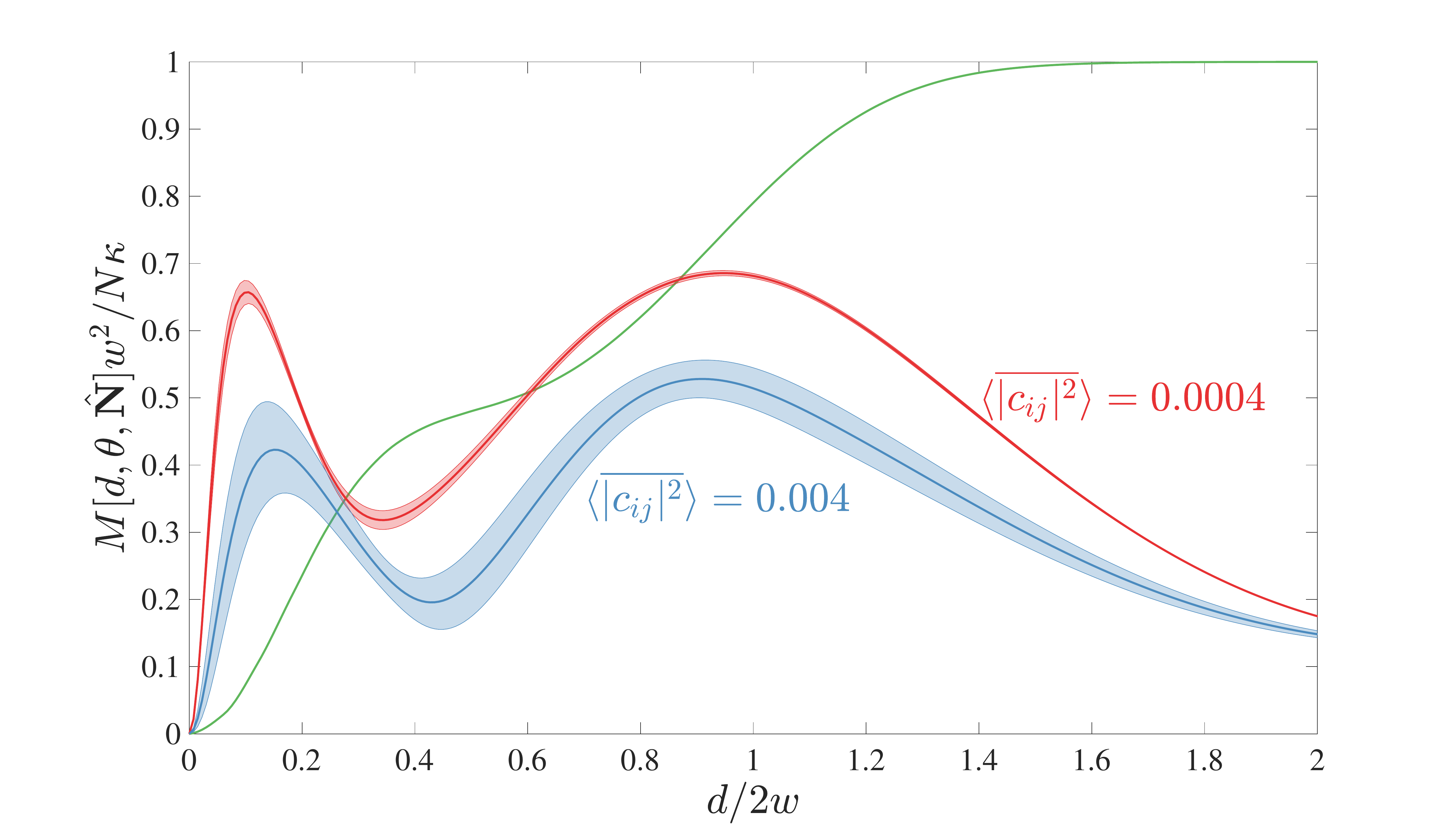}
\caption{Demultiplexing sensitivity $M[d,\theta,\hat{\bf N}]$ in presence of dark counts, $\sigma_k = 0.001 \forall k$, misalignment, $d_s/2w = 0.01$ and $\theta_s = \pi/4$, and two different crosstalk levels, (red) $\langle \overline{|c_{ij}|^2} \rangle = 0.0004$, and (blue) $\langle \overline{|c_{ij}|^2} \rangle = 0.004$.  Red and blue solid lines and bands represent means and standard deviations computed from $500$ realizations of the crosstalk matrices. The green line represents the ideal direct imaging sensitivity. All curves correspond to $\theta = \pi/4$, $N\kappa =10$ and $\gamma=0$.}
\label{Fig:double_cross}
\end{figure}

Despite this loss of sensitivity for small separations, even when all noise sources are considered at the same time, as in Fig.~\ref{Fig:M_zoom} (d), demultiplexing outperforms ideal direct imaging (green lines).
The regime where demultiplexing provides an advantage over direct imaging significantly depends on the brightnesses of the sources, and the noise levels.
In particular, for large mean photon numbers and low noise, there are multiple crossings between the demultiplexing and the ideal direct imaging curves.
To illustrate this behaviour, in Fig.~\ref{Fig:double_cross}, we plot, together with the ideal direct imaging curve (green), two curves corresponding to fixed misalignment and dark count levels but different crosstalk strengths for $N\kappa =10$.
We see that at low crosstalk levels, (red curve in Fig.~\ref{Fig:double_cross}) the demultiplexing sensitivity crosses the ideal direct imaging curve three times. 
Therefore, in the low noise regime, there are regions of larger separations where demultiplexing outperforms direct imaging. 
On the other hand, larger noise levels (blue curve in Fig.~\ref{Fig:double_cross}) reduce the demultiplexing sensitivity and cancel the larger-distance region where demultiplexing outperforms direct imaging.
In Fig.~\ref{main-Fig:diagram} of \cite{sorelli2021optimal}, we investigated how the minimal separation at which the direct imaging sensitivity crosses the demultiplexing sensitivity varies with the noise level for different source brightnesses.

To better understand how the different noise sources affect the sensitivity, we now perform a perturbative expansion of $M[d,\theta, \hat{\bf N}]$ for small separations ($x = d/2w \ll 1$).
For simplicity, we focus on the case of equally bright thermal sources $(\gamma = 0)$.
Moreover, in practically relevant scenarios, we can assume that misalignment is of the order of the separation, or smaller.
Therefore, we assume $x = d/2w \ll 1$ and $x_s = d_s/2w \ll 1$ to be of order $a  = ||(x, x_s)|| \ll 1$ .
From Eq. \eqref{beta}, we then obtain
\begin{subequations}
\begin{align}
\beta_{00}({\bf r}_0 -{\bf r}_s) &= 1-\frac{x^2}{2} -2 x_s^2\nonumber \\& \quad -2 x x_s \cos (\theta -\theta_s) + O\left(a^3\right)\\
\beta_{10}({\bf r}_0 -{\bf r}_s) &=2 x_s \cos\theta _s+x \cos \theta +O\left(a^3\right)\\
\beta_{01}({\bf r}_0 -{\bf r}_s) &=2 x_s \sin\theta _s+x \sin \theta +O\left(a^3\right)\\
\beta_{11}({\bf r}_0 -{\bf r}_s) &=\left(2 x_s \sin \theta _s+x \sin \theta \right)\nonumber\\&\quad \times \left(2 x_s \cos \theta _s+x \cos \theta\right) +O\left(a^3\right)\\
\beta_{20}({\bf r}_0 -{\bf r}_s) &=\frac{\left(2 x_s \cos \theta _s+x \cos \theta\right)^2}{\sqrt{2}}+O\left(a^3\right)\\
\beta_{02}({\bf r}_0 -{\bf r}_s) &=\frac{\left(2 x_s \sin \theta _s+x \sin \theta\right)^2}{\sqrt{2}}+O\left(a^3\right)\\
\beta_{nm}({\bf r}_0 -{\bf r}_s) &= O\left(a^3\right) \quad {n+m\geq 3}.
\end{align}
\label{beta-small}
\end{subequations}
Combining Eqs.~\eqref{f} and \eqref{D_thermal} with the expansions~\eqref{beta-small}, we get the following expression for the derivative vector
\begin{widetext}
\begin{align}
\frac{w}{2N\kappa}D[d,\theta,\hat{\bf N}]_{nm} &= x \left[-2 |c_{nm,00}|^2 + \cos ^2\theta \left(2 \left| c_{nm,10}\right| ^2+\sqrt{2} c_{nm,20} c_{nm,00}^*+\sqrt{2} c_{nm,00} c_{nm,20}^*\right) \right. \nonumber \\
&+ \left. \sin ^2\theta \left(2 \left| c_{nm,01}\right| ^2+\sqrt{2} c_{nm,02} c_{nm,00}^*+\sqrt{2} c_{nm,00} c_{nm,02}^*\right) \right. \\ 
& \left.+\sin (2 \theta ) \left(c_{nm,11} c_{nm,00}^*+c_{nm,10} c_{nm,01}^*+c_{nm,01} c_{nm,10}^*+c_{nm,00} c_{nm,11}^*\right)\right]  + O(a^2), \nonumber
\end{align}
\end{widetext}
where we note that up to first order in $a$, misalignment does not affect the derivative vector.

Let us now consider a generic weak crosstalk distribution, which corresponds to set $\epsilon \ll 1$ in Eq.~\eqref{C}. 
In fact, for $\epsilon \ll 1$, we can write the crosstalk matrix as $C(\epsilon) \approx \mathbf{1} - i \epsilon G$, where the elements of the matrix $G$ are of order unity.
Accordingly, the off diagonal elements of the crosstalk matrix, $c_{nm,kl}$ with $n \neq k$ and $m \neq l$, are of order $\epsilon$ , with $0 < \epsilon \ll 1$.
Under this assumption, we can restrict ourselves to the smallest square covariance matrix containing all terms of order $\epsilon$.
Following this prescription, we obtain the $3 \times 3$ matrix
\begin{equation}
\Gamma^\prime = 
\begin{pmatrix}
\Gamma_{00,00} & \Gamma_{00,01} & \Gamma_{00,10}\\
\Gamma_{01,00} & \Gamma_{01,01} & \Gamma_{01,10}\\
\Gamma_{10,00} & \Gamma_{10,01} & \Gamma_{10,10}
\end{pmatrix} + \begin{pmatrix}
\Sigma  & 0 & 0\\
0 & \Sigma & 0\\
0 & 0 & \Sigma
\end{pmatrix},
\label{Gamma_3x3}
\end{equation} 
with $\Sigma = N^{\rm dc}(N^{\rm dc}+1)$ the dark-count term, which we assumed to be weak, and the same for all modes. 
This truncated covariance matrix contains terms up to order $\epsilon^3$.
In particular, the leading order in $\epsilon$ is, $\epsilon^0$ for $\Gamma_{00,00}$, $\epsilon$ for $\Gamma_{01,00}, \Gamma_{10,00}, \Gamma_{00,01},\Gamma_{00,10}, \Gamma_{10,10}$, and $\Gamma_{01,01}$, and $\epsilon^3$ for $\Gamma_{10,01}$ and $\Gamma_{01,10}$.

The inverse of the $3 \times 3$ matrix in Eq. \eqref{Gamma_3x3} can be obtained analytically and used to determine the measurement sensitivity, which results in
\begin{align}
&M[d,\theta, \hat{\bf N}] = \frac{2 N\kappa}{w^2}  \left(
\frac{\left| c_{00,00}\right|^4}{2 N \kappa (\left| c_{00,00}\right|^4 + \sigma^2)+\left| c_{00,00}\right|^2+\sigma} \right. \nonumber\\ 
&\left. +\frac{\sin ^4\theta \left| c_{01,01}\right|^4+\cos ^4\theta \left| c_{10,10}\right|^4}{2 N\kappa \sigma ^2+\sigma } + O(\epsilon)\right)x^2+O(a^3),
\label{M_sigma0}
\end{align}
where, under the above assumption that detection noise is the same in all modes, we have $\sigma = N^{\rm dc}/2N\kappa$.
In the above expansion, we considered $\sigma$ of order zero in $\epsilon$. 
Accordingly, in Eq.~\eqref{M_sigma0}, the behaviour of the sensitivity for small $x$ is dominated by dark counts (only diagonal terms of the cross-talk matrix appears).

Another relevant scenario might be the one where crosstalk and dark counts give contributions of the same order to the photon counts $N_{nm}$ in a given HG mode, i.e. when $2N\kappa |c_{ij}|^2 \sim N^{\rm dc}$, or in other words when
$\sigma = N^{\rm dc}/2N\kappa$ is of order $\epsilon^2$. 
In this case, the measurement sensitivity would be approximated by 
\begin{align}
M[d,\theta, \hat{\bf N}] &= \frac{2 N\kappa}{w^2}\left(\frac{\sin ^4\theta \left| c_{01,01}\right|^4}{\left| c_{01,00}\right|^2+\sigma} \right.\nonumber\\
&+\left. \frac{\cos ^4\theta\left| c_{10,10}\right|^4}{\left| c_{10,00}\right|^2+\sigma} + O\left(\epsilon^{-1}\right)\right)x^2 + O(a^3).
\label{M_sigma2}
\end{align}
In this case, the probabilities of scattering from the first order modes $u_{01}$ and $u_{10}$ to the fundamental mode $u_{00}$, enters in the behaviour of the measurement sensitivity for small $x$. 
Setting $\sigma = 0$ (corresponding to no dark counts, $N^{\rm dc} =0$) in Eq.~\eqref{M_sigma2}, we obtain the leading order expansion of $M[d,\theta, \hat{\bf N}]$ when only crosstalk is present (see also the Supplementary Material of \cite{gessner2020}).

Equations \eqref{M_sigma0} and \eqref{M_sigma2} can be simplified by considering a uniform crosstalk model \cite{gessner2020}, namely by setting all diagonal entries of the crosstalk matrix to $t$ and all the off-diagonal ones to $r$ such that $|t|^2 +(D-1)|r|^2 =1$. 
For weak crosstalk, we have in addition $|t|^2 \approx 1$ and $|r|^2 \ll 1$. 
Accordingly, Eqs.~\eqref{M_sigma0} and \eqref{M_sigma2} becomes respectively
\begin{align}
M_{\rm dc}[d,\theta, \hat{\bf N}] &\approx \frac{2 N \kappa}{w^2} \left(\frac{\cos (4 \theta )+3}{8 N\kappa \sigma ^2+4 \sigma } \right.  \nonumber \\ 
&\left. \quad +\frac{1}{2 N \kappa \left(\sigma ^2+1\right)+\sigma +1} \right) x^2,\label{M_dc}\\
M_{\rm ct}[d,\theta, \hat{\bf N}] &\approx\frac{2N\kappa}{w^2} \frac{(\cos (4 \theta )+3)}{4 \left(|r|^2+\sigma \right)}x^2. \label{M_ct}
\end{align}
We labelled as $M_{\rm dc}$ ($M_{\rm ct}$) the dark-counts (cross talk) dominated sensitivity.
Equation~\eqref{M_dc} remains valid also in the case when no crosstalk or misalignment are present, and we expand to leading order in $\sigma$.

Interestingly enough, with the assumption that the misalignment is of the same order as the separation, in both Eqs.~\eqref{M_sigma0} and \eqref{M_sigma2} the effect of misalignment is hidden by the other noise sources.
However, in Fig.~\ref{Fig:M_zoom} (a), we know that misalignment alone also cause the measurement sensitivity to go to zero for $x \to 0$.
To illustrate how this happens, we repeat the procedure described above in the absence of crosstalk and dark counts.
The measurement sensitivity, at leading order in $a$, results in
\begin{align}
M_{\rm mis}[d,\theta, \hat{\bf N}] &\approx\frac{2 N\kappa }{w^2} \left(\frac{\sin ^4\theta }{x^2 \sin ^2\theta +4 x_s^2 \sin ^2\theta_s} \right. \nonumber \\ & \left.\quad+\frac{\cos^4\theta }{x^2 \cos ^2\theta +4 x_s^2 \cos ^2\theta_s}\right)x^2.
\label{M_xs}
\end{align}
The approximations of the measurement sensitivity in the different noise regimes described by Eqs.~\eqref{M_sigma0} -- \eqref{M_xs} are compared with the exact numerical results in Fig.~\ref{Fig:M_zoom}.
Interestingly, for all noise sources (Eqs.~\eqref{M_sigma0} -- \eqref{M_xs}) the measurement sensitivity goes to zero as $x^2$ for small separations.
We observed, in Eq.~\eqref{M_di_small}, that also the direct imaging sensitivity (for $\gamma =0$) presents the same scaling: $w^2M_{DI}[d,\theta]/(2N\kappa) \sim  8x^2$.
However, for typical noise levels (as reported in Fig.~\ref{Fig:M_zoom}), demultiplexing has generally a more favourable coefficient.
\begin{table}
\begin{tabular}{c | c c c}
 & Misalignment & Crosstalk & Dark counts \\
\hline
\makecell{Eq.~\eqref{M_sigma0} \\ (Eq.~\eqref{M_dc})}& (\cmark)  & (\cmark, $\epsilon \ll \sigma$) & \cmark\\
\hline 
\makecell{Eq.~\eqref{M_sigma2} \\ (Eq.~\eqref{M_ct})}& (\cmark) & \cmark & (\cmark, $\sigma \lesssim \epsilon^2$)\\
\hline
Eq.~\eqref{M_xs}& \cmark & \xmark & \xmark\\
\hline
\end{tabular}
\caption{Tabular representation of the role of the different noise sources in the perturbative expansions reported in Sec.~\ref{Sec:noisy_sensitivity}: \cmark\; indicates the dominant noise source, (\cmark) indicates a noise source that is present, but not dominant (in case the noise source affect the equation its order is also included in the parenthesis), \xmark\; denotes a noise source that is not included. 
Equation numbers given in parenthesis correspond to the uniform crosstalk model.}
\label{table}
\end{table}

\subsection{Measurement coefficients}
\label{Sec:observable}
\begin{figure}
\centering
\includegraphics[width=\columnwidth]{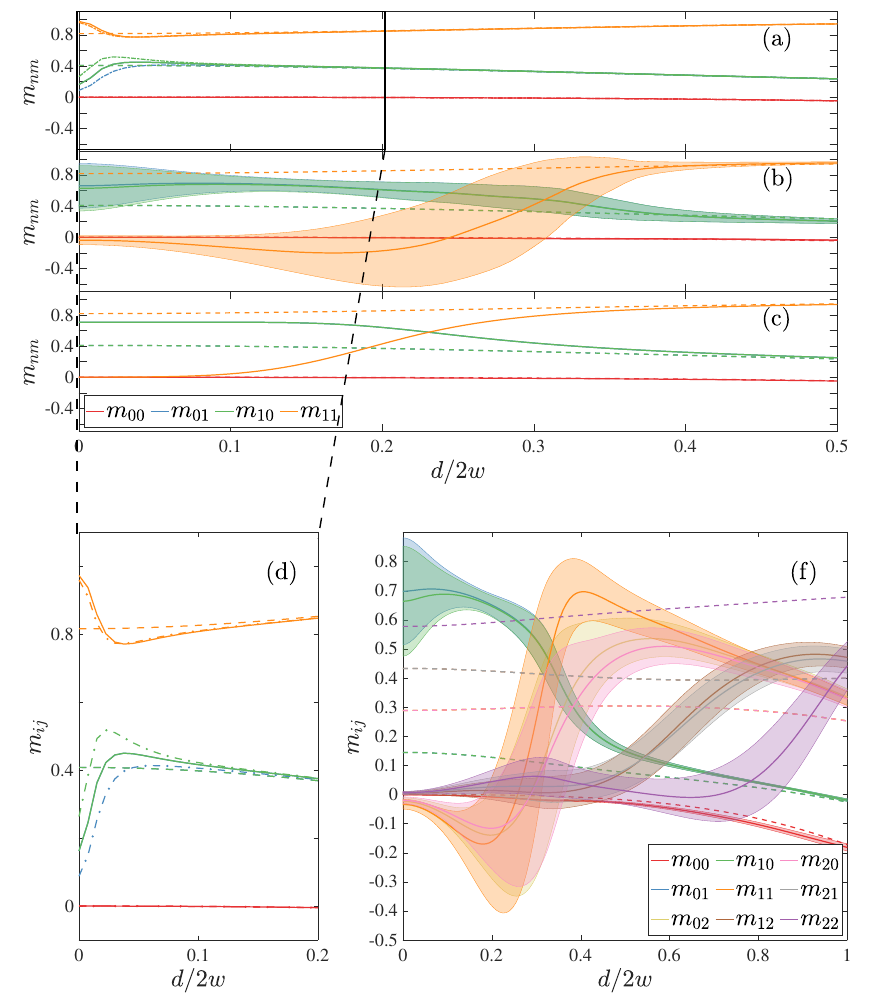}
\caption{Dependence on the source separation $d$ of the measurement coefficients $m_{ij}$ for intensity measurements in the HG modes basis $u_{ij}({\bf r})$ with  (a - d) $0 \leq i, j \leq 1$ ($K=4$) and (f) $0 \leq i, l \leq 2$ ($K=9$).
Different noise sources are considered: (a, d) misalignment ($d_s/2w = 0.01$, $\theta_s = \pi/4$ (solid) and $\theta_s = \pi/3$ (dotted)), (b) crosstalk ($\langle \overline{|c_{ij}|^2} \rangle = 0.0017$), (c) dark counts ($\sigma = 0.001$), and (f) all three combined ($d_s/2w = 0.01$, $\theta_s = \pi/4$, $\langle \overline{|c_{ij}|^2} \rangle = 0.0017$, $\sigma = 0.001$).
Dashed lines represents the coefficients in the noiseless case. Solid lines and bands in (b) and (d) represents the mean and one standard deviation computed over $500$ crosstalk matrices. 
All plots correspond to $N\kappa = 1.5$ and $\theta = \pi/4$.}
\label{Fig:coeff}
\end{figure}
We now discuss the behaviour of the optimal observable that allows to reach the sensitivity bounds illustrated in the previous section.
To this goal, in Fig.~\ref{Fig:coeff}, we study the dependence of the coefficients $m_{ij}$ (see Eq.~\eqref{m}) of the optimal linear combination of intensity measurements $\hat{X}_{\bf m} = \sum_{ij} m_{ij} N_{ij}$ in the HG modes $u_{ij}({\bf r})$ as a function of the separation $d$ between the two sources for $0 \leq i,j \leq Q$ with $Q=1$ (Fig.~\ref{Fig:coeff}  (a-d)) and $Q=2$ (Fig.~\ref{Fig:coeff} (f)). 

Arguably, the most interesting feature of Fig.~\ref{Fig:coeff} is that when $d$ is small compared to the diameter $2w$ of the PSF $u_0({\bf r})$ ($d/2w \lesssim 0.2$), the coefficients $m_{ij}$ depend very weakly on $d$.
Accordingly, in the relevant regime of small separations, a fixed observable can be used to estimate a vast range of parameter values.
An exception to this behaviour is observed in Fig.~\ref{Fig:coeff} (a) and in its zoom, Fig.~\ref{Fig:coeff} (d), where we show the coefficients in presence of misalignment only. 
In this case, we observe that for separations of the order of the misalignment, the measurement coefficients present some modulations.
In particular, for $d \sim d_s$, the shifted centroid significantly impacts  the image decomposition in the $u_{01}({\bf r})$ and $u_{10}({\bf r})$ modes. 
Accordingly, the coefficients for these modes are slightly depleted, while the coefficient of $u_{11}({\bf r})$ mode is increased.
However, when misalignment is combined with other imperfections (Fig.~\ref{Fig:coeff} (f)) these modulations are washed out by the reduction in signal-to-noise ratio induced by crosstalk and dark counts in higher order modes.  

In Sec.~\ref{Sec:Moments_ideal}, we observed that, in the absence of noise, the measurement coefficients $m_{ij}$ only depend on the sum $i+j$ (see Eq.~\eqref{m_th}).  
In Fig.~\ref{Fig:coeff}, we show how different noise sources can break this degeneracy. 
This effect is especially clear for the coefficients $m_{01}$ (blue) and $m_{10}$ (green).
In particular, we see that when the center of the measurement basis does not lie on the sources axis ($\theta_s \neq \theta$, dot-dashed lines in Fig.~\ref{Fig:coeff} (a) and (d)) $m_{01}$ and $m_{10}$ deviate in opposite directions from the $\theta_d =\theta = \pi/4$ curve favouring the mode where the signal increased because of the centroid shift.
Also crosstalk, which affects randomly the different modes, can remove the coefficients' degeneracy, as can be observed in the small difference between the green and blue lines in  Fig.~\ref{Fig:coeff} (b), or in the more evident separation of the yellow and pink curves ($m_{02}$ and $m_{20}$) in Fig.~\ref{Fig:coeff} (f).
On the other hand, in Fig.~\ref{Fig:coeff} (c), where the dark count level was assumed to be the same in all modes, the coefficients $m_{01}$ and $m_{10}$ are perfectly degenerate.
Of course, this would have not been the case, if the electronic noise level had not been the same in the different modes.
Finally, in Fig.~\ref{Fig:coeff} (f), we notice that in the presence of weak noise, even though the degeneracy is removed, the coefficients $m_{ij}$ still move in groups (green-blue for $i+j = 1$,  orange-yellow-pink for $i+j = 2$, brown-grey for $i+j = 3$ and purple for $i+j =4$).

From Fig.~\ref{Fig:coeff} (f), we can also see how the different coefficients change with the separation $d$ between the two sources. 
First of all, for small separations, the mode $u_{00}$ contains no information on $d$, accordingly $m_{00}=0$ for a vast range of separations.
Moreover, in the absence of noise (dashed lines Fig.~\ref{Fig:coeff} (f)), all coefficients $m_{ij>0}$~\eqref{m_th}  are different from zero for all values of $d$, and higher-order modes have larger weights.
In fact, even though, for small separations, the image of the two sources produces very low signals in the higher-order modes, these signals are noiseless.
Accordingly, the optimal observable amplifies these small signals to extract the most information on the parameter $d$ out of them.
On the other hand, different noise sources introduce photon counts that contain no information on the separation.
Accordingly, in the presence of noise (solid lines Fig.~\ref{Fig:coeff} (f)), our optimal observable prescribes to measure only those modes where the signal-to-noise ratio is high enough.
In particular, for small values of $d$, all coefficients are zero (or slightly negative) except $m_{01}$ and $m_{10}$. 
Therefore, demultiplexing into the HG modes $u_{01}({\bf r})$ and $u_{10}({\bf r})$ (blue and green) is sufficient to achieve the optimal resolution.
For larger separations, the other available modes start to become relevant, at first $u_{20}({\bf r})$, $u_{11}({\bf r})$, and $u_{02}({\bf r})$ (orange, yellow and pink), then $u_{12}({\bf r})$, and $u_{21}({\bf r})$ (brown and grey), and finally $u_{22}({\bf r})$ (purple).
Increasing the separation, the higher-order modes become dominant, and all coefficients tend to their values in the absence of noise (the latter effect is visible for $m_{01}$ and $m_{10}$ in Fig.~\ref{Fig:coeff} (f)).

\section{Minimal resolvable distance}
\label{Sec:min_res_dist}
We conclude this work by discussing how the minimal distance, that can be resolved with our moment-based estimation strategy, scales with the number of detected photons. 
For the sake of simplicity, we focus here on thermal sources with equal brightness ($\gamma = 0$).

\begin{figure}
\includegraphics[width=\columnwidth]{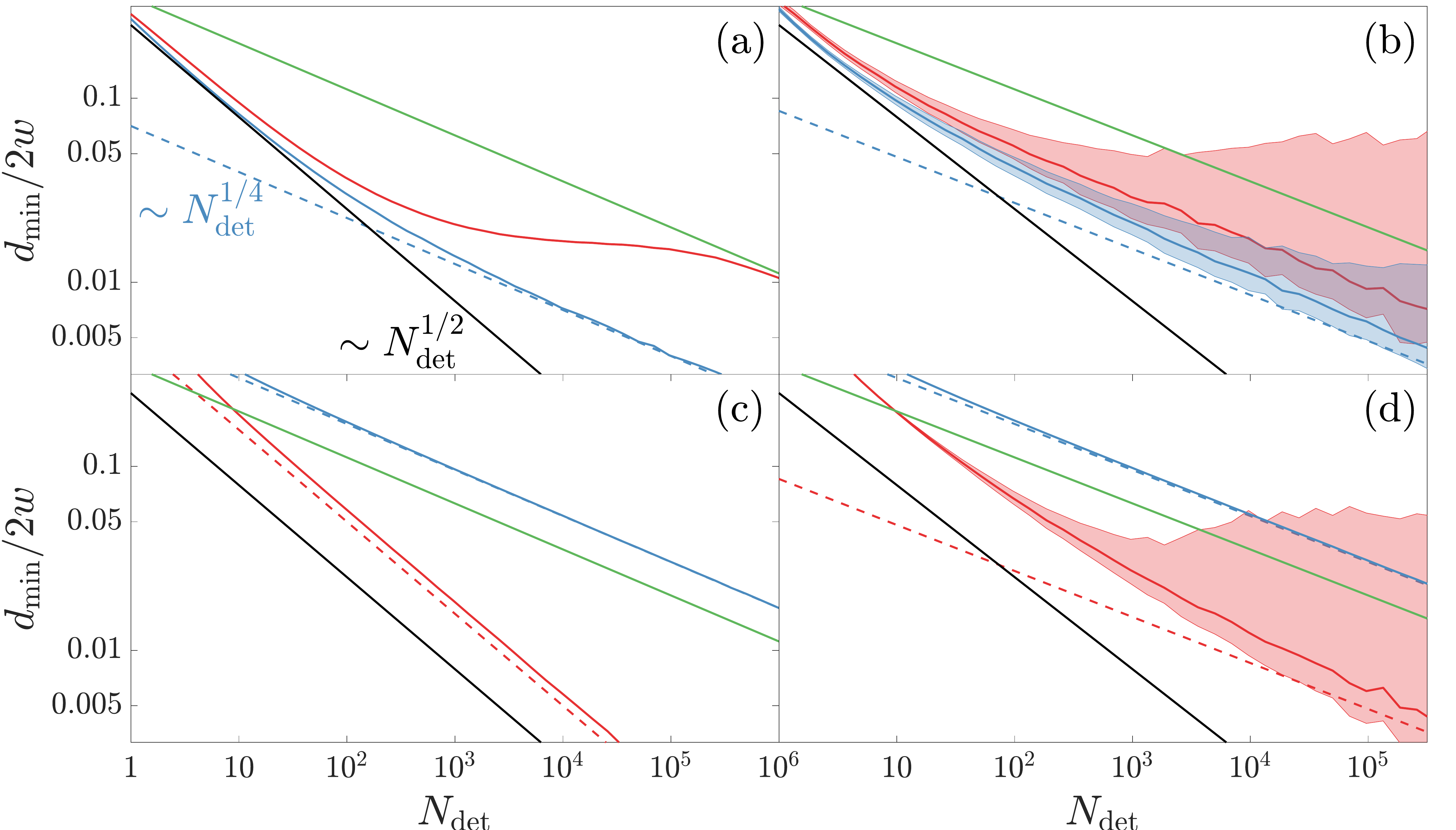}
\caption{Scaling of the minimal resolvable distance $d_{\min}$ with the number of detected photons $N_{\det} = \mu (2 N\kappa)$ changed by varying either the number of measurements $\mu$ (blue) or the brightness of the sources $N$ (red). 
We considered demultiplexing measurements into $K = (Q+1)^2 = 9$ HG modes $u_{ij}({\bf r})$ with $0 \leq i,j \leq Q = 2$ and different noise sources: (a) misalignment ($d_s/2w = 0.01, \theta_s = \pi/4$), (b) crosstalk ($\langle \overline{|c_{ij}|^2} \rangle = 0.0017$), (c) dark counts ($N^{\rm dc} = 1$), and (d) all three combined.
Black and green lines represent the $N_{\det}^{-1/2}$ scaling of ideal demultiplexing, and the $N_{\det}^{-1/4}$ scaling of ideal direct imaging, respectively.
Dashed lines represent analytical approximations for large $N_{\det}$ given by Eqs.~\eqref{d_min_mis} (a), \eqref{d_min_ct} (b) and \eqref{d_min_dc} (c) and (d).
}
\label{Fig:d_min}
\end{figure}

We consider the distance between the two sources to be resolvable as long as the estimation error $\Delta d$ is smaller than the value $d$ of the parameter itself, namely when $d_{\min}/ \Delta d \geq 1$ \cite{gessner2020}.
Following our moment-based estimation approach, we have $(\Delta d)^2 = 1/\mu M[d,\theta, \hat{\bf N}]$, accordingly the minimal resolvable distance is defined by the condition
\begin{equation}
d_{\min} \sqrt{\mu  M[d_{\min},\theta, \hat{\bf N}]} = 1.
\label{d_min}
\end{equation}
Given two thermal sources each emitting on average $N$ photons, an imaging system with a transmissivity $\kappa$, after $\mu$ measurements of our optimal observable, the mean number of detected photons is given by $N_{\det} = \mu 2 N \kappa$ 
\footnote{To be more precise $\mu 2 N \kappa$ is the number of photons received in the image plane, which is always larger equal than the number of detected photons. 
The main cause of undetected photons is the finite number $K$ of modes in the demultiplexing basis. 
However, at the level of the minimal resolvable distance, the population of the high order modes is usually negligible.
It is therefore justified to refer to $N_{\det} = \mu (2 N \kappa)$ as the number of detected photons.}.
Accordingly, for a fixed imaging system, $N_{\det}$ is affected either by $\mu$, which could be changed by increasing or decreasing the detection time, or by the brightness of the sources $N$. 
In Fig.~\ref{Fig:d_min}, we study how for noisy demultiplexing, the solution of Eq.~\eqref{d_min} changes when we vary $N_{\det}$ either by increasing $\mu$ for $N\kappa =1$ (blue curves) or by changing $N\kappa$ for $\mu =1$ (red curves).
While in the first case, every noise source induces the same scaling $d_{\min} \sim N_{\det}^{1/4}$, in the latter case, the non trivial dependence of $M[d_{\min},\theta, \hat{\bf N}]$ on $N$ leads to more complicated behaviours. 
This difference is due to the thermal statistics of our sources.
On the other hand, in the case of coherent sources there would not be any difference, since measuring $\mu$ copies of a coherent state $\ket{\alpha}$ give the same results as measuring $\ket{\mu \alpha}$.

To get an analytical understanding of these behaviours, we use the small-separation expansions of the measurement sensitivity $M[d,\theta,\hat{\bf N}]$.
For ideal demultiplexing, our moment-based estimation strategy saturates the quantum Fisher information~\eqref{M_ideal}. 
In the small $d$ regime, the latter is dominated by the constant term $2N\kappa/w^2$, which defines the solution of Eq.~\eqref{d_min}
\begin{equation}
d_{\min} = \frac{w}{\sqrt{N_{\det}}}.
\label{d_min_id}
\end{equation}
Accordingly, ideal demultiplexing provides a ``shot noise'' scaling of the minimal resolvable distance which is represented as black lines in Fig.~\ref{Fig:d_min}.

Let us now consider the impact of the different noise sources on the minimal resolvable distance achievable with demultiplexing measurements.
For small values of $N_{\det}$, the minimal resolvable distance is determined by the expansion of $M[d,\theta,\hat{\bf N}]$ around a finite value of $d$. 
Independently on the noise level, the leading term of such an expansion is independent on $d$. 
As a consequence, we have the $d_{\min} \sim 1/\sqrt{N_{\det}}$ scaling, we observed in the case of ideal demultiplexing. 
Increasing $N_{\det}$, the solution of Eq.~\eqref{d_min} is determined by the quadratic behaviour of $M[d,\theta, \hat{\bf N}]$ around $d \sim 0$ (see Eqs.~\eqref{M_dc} -\eqref{M_xs}), which induces the changes of scaling observed in Fig.~\ref{Fig:d_min} (similar considerations can be found in \cite{gessner2020}).

To evaluate the impact of misalignment on the large $N_{\det}$ scaling minimal resolvable distance, we use Eq.~\eqref{M_xs}, which we recall is valid for misalignment of the order of the separation $d_s \approx d \ll 1$, and no crosstalk and dark counts (see Table \ref{table}). 
This expression leads to
\begin{equation}
d_{\min} = \frac{\sqrt{2 d_s w}}{N_{\det}^{1/4}(\cos^4\theta \sec^2 \theta_s + \sin^4\theta \csc^2 \theta_s)^{1/4}},
\label{d_min_mis}
\end{equation}
which is represented as a blue dashed line in Fig.~\ref{Fig:d_min} (a).
A similar $\sqrt{d_s}N_{\det}^{-1/4}$ scaling was reported in \cite{AlmeaidaPRA2021} where misalignment was studied in the low brightness regime.
We note that when $N$ is increased for $\mu =1$ the red curve in Fig.~\ref{Fig:d_min} (a) presents an almost flat region before falling on a $N_{\det}^{-1/4}$ line with a less favourable coefficient than the one predicted by Eq.~\eqref{d_min_mis}. 
This different scaling is due to the second order terms in $N\kappa$ which are present in the off-diagonal elements of the covariance matrix~\eqref{Gamma_thermal} whose inverse determines the measurement sensitivity $M[d,\theta,\hat{\bf N}]$.
The analytical expression~\eqref{d_min_mis} cannot capture this behaviour.
In fact, it was derived from Eq.~\eqref{M_xs} which was obtained under the assumption that misalignment is of the same order of the separation $d \sim d_s \sim a$.
The latter condition, which does not hold for the minimal resolvable distance when $N_{\det}$ is large, causes the higher terms in $N\kappa$ to disappear from $M[d,\theta, \hat{\bf N} ]$ when expanding to leading order in $a$. 
On the other hand, for low photon numbers $N\kappa \lesssim 1$, Eq.~\eqref{M_xs} is accurate.
Accordingly  Eq.~\eqref{d_min_mis} perfectly captures the scaling with the detected photon number when $N_{\det}$ is increased by changing the number of measurements $\mu$ with $N\kappa =1$ (blue curve).

Let us now discuss the effect of crosstalk on the minimal resolvable distance. 
To do so, we consider the uniform crosstalk model, and we insert Eq.~\eqref{M_ct} with $\sigma =0$ (no dark counts) into Eq.~\eqref{d_min}, which results in
\begin{equation}
d_{\min} = \frac{w}{N_{\det}^{1/4}}\left( \frac{|r|^2}{3+\cos (4\theta)}\right)^{1/4}.
\label{d_min_ct}
\end{equation}
Equation~\eqref{d_min_ct} is presented as blue dashed lines in Fig.~\eqref{Fig:d_min} (b), and coincides with the results obtained in \cite{gessner2020} for Poissonian sources.
We can see that the blue (fixed $N\kappa =1$ and varying $\mu$) curve approach Eq.~\eqref{d_min_ct} for large $N_d$, while the red (fixed $\mu =1$ and varying $N\kappa$) lies slightly above it.
On the other hand, bright thermal states are more noisy, and this is reflected in the broad red error bands for large values of $N\kappa$.

Finally, let us focus on the role of dark counts. 
In this case, the large $N_{\det}$ scaling of the minimal resolvable distance can be obtained by inserting Eq.~\eqref{M_dc} into Eq.~\eqref{d_min}.
Here, we considered the dark counts level $N^{\rm dc}$  to be constant as we increase $N_{\det}$,
which results in
\begin{align}
d_{\min} &= \frac{\sqrt{2}w}{(N^2\kappa^2\mu)^{1/4}} \left(\frac{\cos (4 \theta )+3}{4N^{\rm dc}(N^{\rm dc} +1)} \right. \nonumber \\
&\quad \left. +\frac{1}{N^{\rm dc}(N^{\rm dc} +1)+ 2N\kappa(2N\kappa +1) }\right)^{-1/4}
\label{d_min_dc}
\end{align}
For large values of $N\kappa$, Eq.~\eqref{d_min_dc} simplifies to 
\begin{equation}
d_{\min} = \frac{\sqrt{2} w}{\sqrt{N\kappa}\mu^{1/4}}\left(\frac{\cos (4 \theta )+3}{N^{\rm dc}(N^{\rm dc}+1)} \right)^{-1/4}.
\label{d_min_dc_largeN}
\end{equation}
From this expression, we can observe that when the number of detected photons $N_{\det}$ is increased by increasing $\mu$, the minimal resolvable distance scales as $d_{\min} \sim N_{\det}^{-1/4}$. 
On the other hand, when $N_{\det}$ is increased with the source brightness, we obtain the same scaling that we had in the ideal case $N_{\det}^{-1/2}$.
This has a very clear physical interpretation: if we detect $N_{\det}$ photons by accumulating several measurements ($\mu \gg 1$) at low brightnesses ($N\kappa \lesssim 1$), we also accumulate dark counts. 
On the other hand, in a single shot measurement ($\mu \gg 1$) at high brightness ($N\kappa \gg 1$) with the same detector dark counts become negligible, we obtain the scaling of an ideal measurement.
These different behaviours can be observed by comparing the scaling of the red ($\mu = 1$, variable $N$) and blue lines ($2N\kappa =1$, variable $\mu$) in Fig.~\ref{Fig:d_min} (c).

Finally, let us mention that even though Eq.~\eqref{d_min_dc_largeN} predicts the correct scaling with $N$ and $\mu$, it tends to underestimate $d_{\min}$. 
This is due to the quadratic approximation \eqref{M_dc} which, as visible in Fig.~\ref{Fig:M_zoom} (c), grows faster than the exact measurement sensitivity, and consequently $d\sqrt{\mu M_{\rm dc}[d,\theta,\hat{\bf N}]}$ crosses $1$ (see Eq.~\eqref{d_min}) earlier than the exact curve  $d\sqrt{\mu M[d,\theta,\hat{\bf N}]}$. 
This underestimation is particularly relevant for low dark counts levels, $N^{\rm dc} \ll 1$, and in particular it leads to an unphysical better-than-ideal scaling for $N^{\rm dc} < (\sqrt{4+2/\mu}-2)/4$.

In Fig.~\ref{Fig:d_min} (d), we present the behaviour of the minimal resolvable distance, when all noise sources are present at the same time. 
As discussed above, when the number of detected photons $N_{\det}$ is increased by accumulating many measurements $\mu$ (blue curves), we are also increasing the dark counts.
In this case, dark counts are the dominant noise source (compare with the blue curve in Fig.~\ref{Fig:d_min} (c)), and the error bands due to crosstalk become negligible. 
On the other hand, when $N_{\det}$ is increased with the brightness of the sources $N$ (red curve in Fig.~\ref{Fig:d_min} (d)), dark counts quickly become negligible, and the dominant noise source is crosstalk (compare with the red curve in Fig.~\ref{Fig:d_min} (c)).

In conclusion, let us compare the noisy demultiplexing results with those of direct imaging. 
The  large $N_d$ scaling of the minimal resolvable distance for noiseless direct imaging can be obtained from Eq.~\eqref{M_di_small}, which, for $\gamma = 0$, results in $M[d,\theta,N_p] \approx 8 x^2$, and leads to (see also \cite{gessner2020})
\begin{equation}
d_{\min} = \frac{w}{(N_d)^{1/4}}\left(\frac{1}{2} \right)^{1/4}.
\label{d_min_di}
\end{equation}
Equation~\eqref{d_min_di} is plotted as green lines in Fig.~\ref{Fig:d_min} and for a vast range of noise conditions it stays above the noisy demultiplexing curves.
An exception is represented by the blue curves in Fig.~\ref{Fig:d_min} (c) and (d). 
However, in Fig.~\ref{Fig:d_min} (c) and (d), we chose $N^{\rm dc} =1$ for which the approximation \eqref{d_min_dc_largeN} gives sensible results. 
For large source brightnesses, this value corresponds to a reasonable ratio between dark counts and photons counts in each demultiplexed mode.
On the other hand, it is overpessimistic for $2N\kappa = 1$, as it can be quickly understood by considering, for example, that $N_{01}/2N\kappa \approx 10^{-5}$ at $x = d/2w \approx 10^{-2}$.
Therefore, we can conclude that, in most practical situations, spatial mode demultiplexing allows to resolve significantly smaller separations than direct imaging.

\section{Conclusion}
\label{Sec:conclusion}
We discussed in detail a method to extract the separation of two thermal sources, with arbitrary and possibly different brightnesses, from a single optimized measurement observable. 
For this imaging application, we considered different measurements such as direct imaging with pixel detectors and realistic spatial mode demultiplexing.
Our results show how, even in presence of different relevant noise sources, demultiplexing allows for better resolutions than direct imaging.
In several realistic scenarios, we constructed the optimal observable for demultiplexing measurements, which depends very weakly on the separation $d$, providing stability of the estimation procedure over a vast domain of separations.
Moreover, in the limiting case of noiseless demultiplexing, for arbitrary received photon numbers and separations, the optimized observable approaches the quantum Cram\'er-Rao bound if sufficiently many modes are measured.
Finally, for low photon numbers in the image plane, the sensitivity $M[d,\theta,\hat{\bf N}]$ of our method saturates the Fisher information even for noisy demultiplexing.

From our results is evident that the coefficients of the optimal linear combination of photon number measurements are severely affected by noise. 
This dependence must be taken into account if one wants to achieve optimality.
This could be done by calculating the coefficients from experimental measurements of the covariance matrix and the derivative vector, obtained with the help of two test sources. 
Alternatively, one could characterize the different noise sources, and then compute the coefficients theoretically.
In this latter case, a precise measure of the crosstalk matrix of a demultiplexer, and the dark count level of detectors are not problematic. 
On the other hand, misalignment errors are mainly due to an imperfect knowledge of the source centroid.
These errors may be limited by scanning the multiplexer position, or by using adaptive methods such as the one proposed in \cite{Grace:20}.

Finally, let us point out that in experimental realizations (see for example \cite{Boucher:20}) all available observables can be measured at the same time which allows to look at arbitrary linear combinations in post processing.
Therefore, the estimation technique presented here can be understood as the optimal post-processing technique that makes the best use of the available data.
As a consequence, our approach fits particularly well for the estimation of a dynamically changing parameter. 
In fact, provided that the photon detectors are faster than the typical time scale of the parameter changes, the optimal measurement coefficients at each time can be selected in post-processing.

\acknowledgments
GS acknowledges financial support of ONERA - the French aerospace lab.  MG acknowledges funding by the LabEx ENS-ICFP:ANR-10-LABX-0010/ANR-10-IDEX-0001-02 PSL*. This work was partially funded by French ANR under COSMIC project (ANR-19-ASTR-0020-01). This work received funding from the European Union’s Horizon 2020 research and innovation programme under grant agreement No 899587. This work was supported by the European Union’s Horizon 2020 research and innovation programme under the QuantERA programme through the project ApresSF.
\bibliography{SR}{}
\end{document}